\documentclass[aps,twocolumn,showpacs,preprintnumbers,amsmath,amssymb]{revtex4-1} 
\usepackage{dcolumn}
\usepackage{bm}

\usepackage{amsmath}
\usepackage{amssymb}
\usepackage{amsfonts} 
\usepackage{wasysym}  
\usepackage{graphics}  
\usepackage{psfrag} 
\usepackage{graphicx} 
\usepackage{epsfig}    
\usepackage{rotating}  
 \usepackage[latin1]{inputenc}
 \usepackage{color}
 \usepackage{rotating}
\usepackage{csquotes}
 \usepackage{multirow}
\usepackage[nooneline,tight,raggedright,FIGTOPCAP]{subfigure}
\usepackage[linkcolor=red,citecolor=blue,urlcolor=blue,colorlinks=true]{hyperref}
\usepackage{nameref,hyperref}

\usepackage{siunitx} 

\newcolumntype{+}{!{\vrule width 2pt}}

\newlength\savedwidth

\newcommand\thickhline{\noalign{\global\savedwidth\arrayrulewidth\global\arrayrulewidth 2pt}%
\hline
\noalign{\global\arrayrulewidth\savedwidth}}

\begin{document}

\title{Regulation of Pom cluster dynamics in \textit{Myxococcus xanthus}}

\author{Silke Bergeler}
\author{Erwin Frey}
 \email{frey@lmu.de}
\affiliation{
 Arnold-Sommerfeld-Center for Theoretical Physics and Center for NanoScience, Department of Physics, Ludwig-Maximilians-Universit{\"a}t M{\"u}nchen, Munich, Germany
}


\begin{abstract}
Precise positioning of the cell division site is essential for the correct segregation of the genetic material into the two daughter cells. In the bacterium \textit{Myxococcus xanthus}, the proteins PomX and PomY form a cluster on the chromosome that performs a biased random walk to midcell and positively regulates cell division there. PomZ, an ATPase, is necessary for tethering of the cluster to the nucleoid and regulates its movement towards midcell. 
It has remained unclear how the cluster dynamics change when the biochemical parameters, such as the attachment rates of PomZ to the nucleoid and the cluster, the ATP hydrolysis rate of PomZ or the mobility of PomZ dimers interacting with the nucleoid and cluster, are varied. 
To answer these questions, we investigate a one-dimensional model that includes the nucleoid, the Pom cluster and the PomZ protein. We find that a mechanism based on the diffusive PomZ fluxes on the nucleoid into the cluster can explain the latter's midnucleoid localization for a broad parameter range. 
Furthermore, there is an ATP hydrolysis rate that minimizes the time the cluster needs to reach midnucleoid. If the dynamics of PomZ dimers on the nucleoid is slow relative to the cluster's velocity, we observe oscillatory cluster movements around midnucleoid. To understand midnucleoid localization, we developed a semi-analytical approach that dissects the net movement of the cluster into its components: the difference in PomZ fluxes into the cluster from either side, the force exerted by a single PomZ dimer on the cluster and the effective friction coefficient of the cluster. 
Importantly, we predict that the Pom cluster oscillates around midnucleoid if the diffusivity of PomZ on the nucleoid is reduced. A similar approach to that applied here may also prove useful for cargo localization in ParAB\textit{S} systems.  
\end{abstract}

\maketitle

\section*{Author summary}
In order for the rod-shaped bacterium \textit{M. xanthus} to reproduce, its genetic content must be duplicated, distributed equally to the two cell halves and then the cell must divide precisely at midcell. Three proteins, called PomX, PomY and PomZ, are important for the localization of the cell division site at midcell. PomX and PomY form a cluster and PomZ tethers this cluster to the bacterial DNA or nucleoid (complex of chromosomal DNA and proteins) and is important for the movement of the cluster from the nucleoid pole towards midcell. We are interested in the question how the cluster trajectories change when the PomZ dynamics is varied. To investigate this question we developed a mathematical model that incorporates the nucleoid, the cluster and the PomZ dimers. We simulated the cluster trajectories for different model parameters, such as different diffusion constants of PomZ on the nucleoid. Interestingly, when the PomZ dimers diffuse slowly on the nucleoid, we observed oscillatory cluster movements around midcell. Our results provide insights into intracellular positioning of proteins generally. 

\section*{Introduction}
The formation of protein patterns and the intracellular positioning of proteins is a major prerequisite for many important processes in bacterial cells, such as cell division. In order to maintain the genetic content of the bacterial cell, the chromosome (nucleoid) is duplicated during the cell cycle and must be segregated into the two cell halves prior to cell division. The future division site is defined by the FtsZ ring, which forms at midcell and recruits the cytokinetic machinery. Interestingly, FtsZ is highly conserved in bacteria, but the protein systems responsible for the positioning of the FtsZ ring, and with it the cell division site, are not \cite{Rothfield2005, Thanbichler2006, Lutkenhaus2012}. 

Recently, Schumacher et al.\ identified a set of proteins, called PomX, PomY and PomZ, in \textit{Myxococcus xanthus} cells that are important for midcell localization and formation of the FtsZ ring \cite{Treuner-Lange2013, Schumacher2017a, Schumacher2017}. PomZ is an ATPase, which belongs to the family of ParA/MinD ATPases. It binds non-specifically to DNA in its dimeric, ATP-bound state, and its activity is stimulated by interactions with PomX, PomY and DNA. PomX and PomY form a single cluster, which is tethered to the nucleoid via PomZ dimers bound to the chromosome. Starting from an off-center position near one nucleoid pole, the cluster moves towards midnucleoid, coinciding with midcell \cite{Schumacher2017}. When the cluster has reached midcell, the FtsZ ring forms there and the cell divides. During cell division, the cluster splits into two halves, such that each half is located at one pole of the nucleoids in the daughter cells, and the same cycle repeats. Notably, the Pom proteins localize to midcell before FtsZ and also in the absence of FtsZ \cite{Treuner-Lange2013, Schumacher2017a}.

Midcell localization of the FtsZ ring has been well studied in \textit{Escherichia coli} cells \cite{Howard2001, Kruse2002, Huang2003, Fange2006, Touhami2006, Lutkenhaus2007, Loose2008, Halatek2012, Halatek2014a, Wu2016, Thalmeier2016, Halatek2018, Frey2018}. Here, Min proteins (MinC, MinD and MinE) guide the formation of the FtsZ ring at midcell. Both systems contain an ATPase (PomZ and MinD, respectively) and perform the same task in the cell, i.e.\ midcell sensing. Nevertheless, the two systems differ in various ways: First, the scaffold to which the ATP-bound ATPase binds is different: MinD binds to the cell membrane and PomZ to the bacterial nucleoid in the cytoplasm. Second, MinD-bound MinC inhibits \cite{Hu1999}, whereas the Pom cluster promotes FtsZ ring formation at midcell \cite{Schumacher2017a}. Finally, the observed protein patterns differ: the Pom proteins colocalize in a cluster that moves towards midcell, while the Min proteins, which do not form a cluster, oscillate from pole to pole \cite{Raskin1999, Hu1999}. The oscillatory pole-to-pole movement of the Min proteins results in a minimal Min protein concentration at midcell on average over time. Since FtsZ ring formation is negatively regulated by MinC, this restricts the ring to midcell. From a mechanistic point of view, the Pom system is closer to plasmid and chromosome segregation systems that involve a ParAB\textit{S} system than to the Min system.

Like the Pom system, plasmid and chromosome segregation systems make use of an ATPase that shuttles one or several cargoes (such as a plasmid, a partition complex or a protein cluster) along the nucleoid. Low-copy-number plasmids need to be actively segregated to ensure that both daughter cells inherit a copy of the plasmid. To ensure their equal distribution to the daughter cells, the plasmids are tethered to the nucleoid and positioned at equal distances along the nucleoid by ParAB\textit{S} systems \cite{Ringgaard2009, Ebersbach2006, Sengupta2010}. A ParAB\textit{S} system consists of the proteins ParA and ParB, and a DNA sequence, \textit{parS}. ParA proteins are ATPases, which bind non-specifically to DNA as ATP-bound dimers \cite{Leonard2005, Hester2007, Scholefield2011}. Their ATPase activity is stimulated in the presence of ParB \cite{Ptacin2010, Schofield2010, Vecchiarelli2013}, which binds to the \textit{parS} sequence on the chromosome (to form the partition complex) or on the plasmid \cite{Lutkenhaus2012}. Besides plasmid and chromosome segregation \cite{Gerdes2010}, ParAB\textit{S} systems are also important for the positioning of cellular components (e.g. chemotactic clusters or carboxysomes) \cite{Roberts2012, Savage2010}. Several different cargo dynamics involving ParAB\textit{S} systems have been observed: oscillatory movement of ParA and its cargo \cite{Hatano2007, Ringgaard2009}, equidistant positioning of multiple cargoes \cite{Ringgaard2009, Ebersbach2006, Sengupta2010} and movement from one cell pole to the other \cite{Lim2014}.

To account for the dynamics observed in Par systems, various mechanisms have been proposed. Some models rely on ParA filament formation \cite{Ringgaard2009, Ptacin2010, Howard2010, Fogel2006, Banigan2011, Shtylla2012, Gerdes2010}; others challenge this assumption in \textit{in vivo} systems \cite{LeGall2016, Lim2014, Vecchiarelli2010}. A diffusion-ratchet mechanism for the movement of ParB-coated beads \textit{in vitro} and DNA segregation \textit{in vivo} has been introduced \cite{Vecchiarelli2010, Vecchiarelli2013, Hwang2013, Vecchiarelli2014, Hu2017a, Hu2017}. Based on the observation that DNA has elastic properties \cite{Wiggins2010, Lim2014}, a DNA relay mechanism for the movement of the partition complex was proposed \cite{Lim2014, Surovtsev2016a}. Here, the force exerted on the cargo is attributed to the elastic properties of the chromosome. Ietswaart et al.\ observed that if a plasmid is located off-center on the nucleoid, the ParA flux from the left and right sides of the plasmid differ \cite{Ietswaart2014}. Based on this idea, they proposed a model that produces equal plasmid spacing over the nucleoid as long as the plasmid moves in the direction of the higher ParA concentration \cite{Ietswaart2014}. Additionally, models based on reaction-diffusion equations for Par protein dynamics, have been introduced \cite{Murray2017, Walter2017, Sugawara2011, Jindal2015}. 

The Pom system in \textit{M. xanthus} cells differs from the ParAB\textit{S} systems described so far by its large cargo (Pom cluster) size, a high PomZ density in direct association with the cluster and distinctive cargo dynamics: In wild-type \textit{M. xanthus} cells, the Pom cluster is positioned at midcell and no pole-to-pole oscillations of the cluster on the nucleoid are observed \cite{Schumacher2017a}. In order to account for the experimental observations in \textit{M. xanthus} cells, we have proposed a model for midcell localization \cite{Schumacher2017a} that includes the elasticity of the nucleoid and the PomZ proteins \cite{Lim2014, Lansky2015}. Our model suggests a positioning mechanism that relies on the biasing of fluxes of PomZ dimers on the nucleoid, similar to the equipositioning mechanism proposed by Ietswaart et al. \cite{Ietswaart2014}. With this model we were able to reproduce midnucleoid localization with physiologically relevant parameters \cite{Schumacher2017a}, but it remained unclear how the movement of the cluster changes when the rates of the key biological processes involved are varied.

Here, we investigate Pom cluster dynamics when the attachment rate of PomZ dimers to the nucleoid, the binding rate of PomZ dimers to the cluster, the ATP hydrolysis rate of PomZ dimers and the diffusivity of PomZ dimers on the nucleoid and cluster are each varied over a broad parameter range. Interestingly, we observed that there exists an intermediate ATP hydrolysis rate that minimizes the time the clusters need to reach midnucleoid. Furthermore, we found that fast diffusion of PomZ dimers on the cluster accelerates the movement of the cluster towards midnucleoid. To gain a better understanding of the cluster dynamics observed in the \textit{in-silico} parameter sweeps, we investigated how PomZ dimers generate a net force on the cluster in our model. For the case where the PomZ gradient builds up faster than the velocity of cluster movement, we derived a semi-analytical approximation for the average cluster trajectory, which dissects the generation of a net force into two parts: the difference between the diffusive PomZ fluxes into the cluster from either side, and the force exerted by a single PomZ dimer during its interaction with the cluster. This net force can account for the movement of the cluster to midnucleoid. In contrast, when the PomZ dimers diffuse slowly on the nucleoid, we observed oscillatory cluster movement.  

\begin{figure*}[t]
\centering
\includegraphics[]{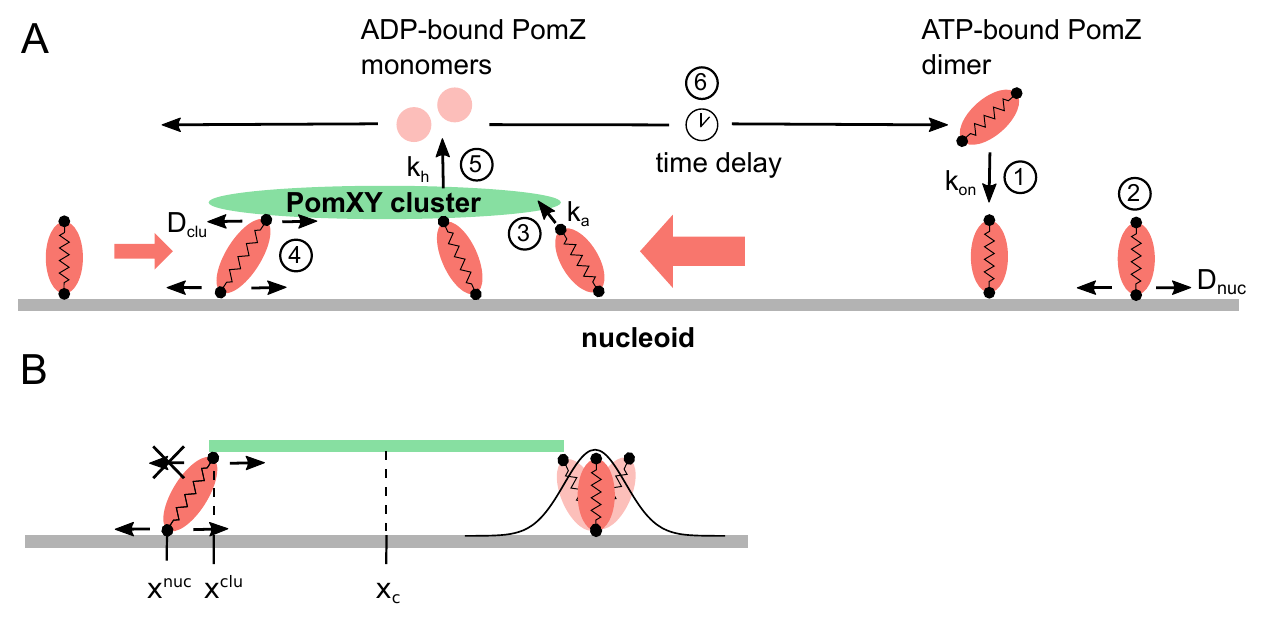}
\caption{\label{fig:stochastic_model} \textbf{Flux-based model for midnucleoid positioning.} (A) In our mathematical model, ATP-bound PomZ dimers can attach to the nucleoid (1) and then diffuse along it (2). The elasticity of the chromosome and the PomZ dimers is effectively included by modelling the PomZ dimers as springs. A nucleoid-bound PomZ dimer has a free binding site available to bind to the PomXY cluster (3). When also bound to the PomXY cluster, a PomZ dimer can diffuse both on the cluster and on the nucleoid (4). The interaction of PomZ with the PomXY cluster (and DNA) leads to a stimulation of the ATPase activity of PomZ, which in turn causes a conformational change in the ATP-bound PomZ dimer and the release of two ADP-bound monomers into the cytosol (5). ADP-bound PomZ monomers must exchange ADP for ATP and form dimers before they can bind to the nucleoid again. Hence, there is a delay between release of the inactive, ADP-bound form and reconstitution of the active, ATP-bound form (6). (B) Details of the PomZ interactions with the PomXY cluster. Not only PomZ dimers with a nucleoid binding site below the PomXY cluster, but also PomZ dimers outside of the cluster region can attach to the cluster, in a stretched configuration. The edges of the PomXY cluster are reflecting boundary conditions for the movement of PomZ's cluster binding site (indicated by the crossed arrow).}
\end{figure*}

\section*{Results}

\subsection*{Stochastic model}
Previously, we developed a stochastic lattice gas model to understand the dynamics of the PomXY cluster, i.e.\ the cluster consisting of PomX and PomY proteins, in \textit{M. xanthus} bacterial cells \cite{Schumacher2017a}. In this model, both the nucleoid and the PomXY cluster are reduced to one-dimensional lattices of length $L$ and $L_c$, respectively (Fig.~\ref{fig:stochastic_model}). Note, that we regard the PomXY cluster composition as static in our model. The PomZ dimer dynamics is modelled as follows: ATP-bound PomZ dimers can bind to the nucleoid with rate $k_\text{on}$ (Fig.~\ref{fig:stochastic_model}A(1)), except where the PomXY cluster is located, and diffuse on the nucleoid with diffusion coefficient $D_\text{nuc}$ (Fig.~\ref{fig:stochastic_model}A(2)). We model the PomZ dimers effectively as springs with spring stiffness $k$ to account for the elastic properties of the chromosome and the PomZ dimers. We expect the PomZ dimers to be stiffer than the chromosome, such that the elasticity of the nucleoid is the main contribution (see also \cite{Lim2014}, \cite{Hu2017a}). A PomZ dimer has two binding sites, and attaches to the nucleoid via the first. Because of thermal fluctuations, the relative position of the second binding site, which enables PomZ to bind to the PomXY cluster, is distributed according to a Boltzmann distribution with the energy of the spring. Therefore, PomZ dimers can attach to the PomXY cluster even if their nucleoid binding sites are not directly below the cluster (Fig.~\ref{fig:stochastic_model}A(3) and \ref{fig:stochastic_model}B). We include this factor in the model by multiplying the rate of attachment, $k_a^0$, by the Boltzmann factor corresponding to the energy of the spring (as in \cite{Lansky2015}):
\begin{equation}
k_a = k_a^0 \exp\left[-\frac{1}{2} \frac{k}{k_B T} (x_i^\text{clu} - x_i^\text{nuc})^2 \right].
\end{equation}
The positions of the cluster and nucleoid binding sites of the $i$-th PomZ dimer bound to the nucleoid and the PomXY cluster are given as $x_i^\text{clu}$ and $x_i^\text{nuc}$, respectively (see Fig.~\ref{fig:stochastic_model}B). 

PomZ dimers bound to the PomXY cluster and the nucleoid are assumed to diffuse on both scaffolds (Fig.~\ref{fig:stochastic_model}A(4)). This assumption is motivated by the experimental observation that flourescently tagged PomZ brightly stains the whole cluster, and not just the cluster boundary, in fluorescence micrographs \cite{Schumacher2017a}. We allow for different diffusivities of the PomZ dimers on the PomXY cluster and the nucleoid. The hopping rates of the nucleoid and cluster binding sites are $\epsilon_\text{hop, nuc}^0 = D_\text{nuc}/a^2$ and $\epsilon_\text{hop, clu}^0 = D_\text{clu}/a^2$, with the lattice spacing $a$, respectively, being weighted by a Boltzmann factor that accounts for the energy change of the spring due to the movement:
\begin{align}
\label{eq:hopping_rates}
\epsilon_\text{hop} & = \epsilon_\text{hop}^0  \\
\nonumber
& \times e^{-\frac{1}{4} \frac{k}{k_B T} \left((x_i^\text{clu, to} - x_i^\text{nuc, to})^2 - (x_i^\text{clu, from} - x_i^\text{nuc, from})^2\right)}.
\end{align}
Here, $x_i^\text{clu, from}, x_i^\text{nuc, from}$ and $x_i^\text{clu, to}, x_i^\text{nuc, to}$ signify the position of the binding sites of the $i$-th PomZ dimer to the cluster and nucleoid before and after hopping, respectively. The additional factor of $1/2$ in the exponent is chosen such that detailed balance holds for the hopping events and the rates for hopping to a neighboring site and hopping back are the inverse of each other (see \cite{Lansky2015}). Because of the exponential factor in Eq.~\ref{eq:hopping_rates} a PomZ dimer is most likely to move in the direction that relaxes the spring. We chose reflecting boundary conditions for diffusion of PomZ on both the nucleoid and the PomXY cluster. PomX, PomY and DNA stimulate the ATPase activity of PomZ, which leads to a conformational change and finally to detachment of two ADP-bound PomZ monomers from the nucleoid \cite{Schumacher2017a}. In our model, we combine the processes of nucleotide hydrolysis and detachment into one rate by assuming that nucleoid- and cluster-bound PomZ dimers are released into the cytosol with hydrolysis rate $k_h$ (Fig.~\ref{fig:stochastic_model}A(5)). The ADP-bound PomZ monomers must then exchange ADP for ATP and dimerize before they can rebind to the nucleoid. This leads to a delay between detachment from and reattachment to the nucleoid (Fig.~\ref{fig:stochastic_model}A(6)). The delay and the rapid diffusion of PomZ in the cytosol \cite{Schumacher2017a} lead to an essentially homogeneous distribution of ATP-bound, dimeric PomZ in the cytosol. Hence, our assumption of a homogeneous attachment rate of cytosolic ATP-bound PomZ dimers to the nucleoid is justified. Note that we do not include ADP-bound PomZ and PomZ monomers explicitly in our model, but only consider PomZ proteins in the ATP-bound dimeric state. The total number of PomZ dimers is assumed to be constant and is denoted by $N_\text{total}$.

So far we have described the stochastic dynamics of the PomZ dimers. The interactions of PomZ dimers with the PomXY cluster result in forces being exerted on the cluster, which cause it to move on the nucleoid. The observable of interest is the PomXY cluster position, $x_c$, over time. We approximate the PomXY cluster dynamics as overdamped, such that the equation of motion for the cluster position is given by the following force balance equation
\begin{equation}
\gamma_c \partial_t x_c = -k \sum_{i=1}^N (x_i^\text{clu}-x_i^\text{nuc}),
\label{eq:eom_cluster}
\end{equation}
with $\gamma_c$ being the friction coefficient of the PomXY cluster in the cytosol and $N$ the total number of cluster-bound PomZ dimers. We focus on the stochasticity in the interactions of PomZ dimers with the PomXY cluster, which in turn lead to stochastic forces acting on the cluster, and disregard movements of the cluster due to thermal noise. Therefore, we do not include a Langevin noise term in Eq.~\ref{eq:eom_cluster}. 

\subsection*{\textit{In-silico} parameter analysis}
\begin{figure*}[t]
\includegraphics[]{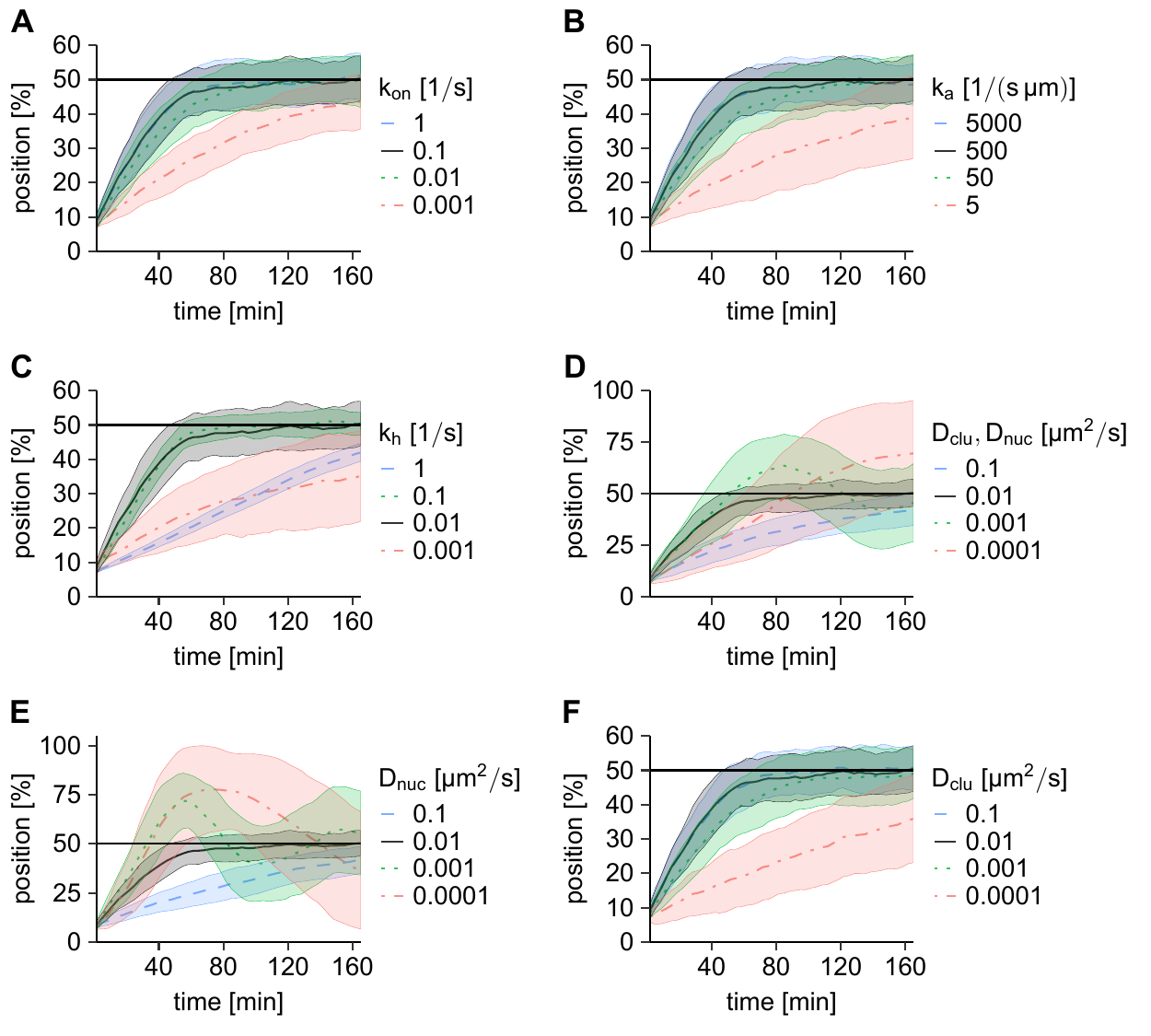}
\caption{\textbf{Exploring the parameter space.} (A-F) Stochastic simulations show different qualitative behavior of the PomXY cluster trajectories when the model parameters are altered. We performed stochastic simulations using the parameter set given in Table \ref{S1_Table}, with one of the parameters varied as indicated. In D, the diffusion constants of PomZ on the nucleoid and on the PomXY cluster are set to the same value. The result for the parameter set given in Table \ref{S1_Table} is always shown in black for comparison purposes. The average cluster trajectories are shown as unbroken or dashed lines and the shaded regions indicate the region of $\pm$ one standard deviation. In the simulations, the initial position of the PomXY cluster is chosen such that the left edge of the cluster coincides with the left edge of the nucleoid (for more details see Materials and methods section). For the calculation of the mean and standard deviations the cluster positions are grouped into $50$ time intervals. For each parameter set we simulated $100$ trajectories.}
\label{fig:parameter_sweep}
\end{figure*}

With physiologically relevant parameters, either measured or taken from the literature on related systems (Table \ref{S1_Table}, data shown in black), the stochastic simulations show midnucleoid positioning of the PomXY cluster (Fig.~\ref{fig:parameter_sweep}). The underlying mechanism for midnucleoid localization is based on the flux of PomZ on the nucleoid, which can be described as follows. If the PomXY cluster is located to the left of midnucleoid, the average flux of PomZ dimers into the cluster from the right is larger than that from the left \cite{Ietswaart2014} (red arrows in Fig.~\ref{fig:stochastic_model}A). Particles that attach to the cluster typically exert a net force in the direction from which they attach. Thus the flux imbalance leads to a net force towards the right, i.e.\ towards midnucleoid. If the cluster overshoots midnucleoid or is already positioned to the right of midnucleoid, the asymmetry in the fluxes is reversed and the cluster moves back towards midnucleoid. Overall, this leads to a self-regulating process that positions the PomXY cluster at midnucleoid. 

Wild-type cells contain a total PomZ dimer number of $N_\text{total} \approx 100$ \cite{Schumacher2017a}, which results in a low density of PomZ dimers on the nucleoid, and hence exclusion effects can be neglected. Here, we focus on the low PomZ density regime, which reflects the wild-type situation, and therefore we do not limit the PomZ dimer density on the nucleoid. We performed parameter sweeps by varying the attachment rate of PomZ dimers to the nucleoid, $k_\text{on}$, the binding rate of nucleoid-bound PomZ dimers to the PomXY cluster, $k_a$, the ATP hydrolysis rate of PomZ dimers, $k_h$, and the mobility of PomZ dimers on the nucleoid, $D_\text{nuc}$, and on the PomXY cluster, $D_\text{clu}$, over a broad range (Fig.~\ref{fig:parameter_sweep}). 

The parameter sweeps show that increasing the attachment rate to the nucleoid, $k_\text{on}$, or the binding rate to the PomXY cluster, $k_a$, decreases the time the cluster needs to reach midnucleoid (Fig.~\ref{fig:parameter_sweep}A and \ref{fig:parameter_sweep}B). In both cases, the trajectories become independent of the particular parameter when its value exceeds a certain threshold. We conclude that increasing the rate of attachment of PomZ to the nucleoid or the binding of PomZ to the PomXY cluster speeds up the positioning process until an optimum is reached. 

Next, we consider the effects of varying the rate of ATP hydrolysis by PomZ dimers associated with the PomXY cluster, which is important to maintain the cyclic flux of PomZ dimers between cytosolic and DNA-bound states. This rate also sets the time scale for the interaction of PomZ dimers with the PomXY cluster. 
The simulations show that decreasing the hydrolysis rate ($k_h=\SI{0.001}{s^{-1}}$) reduces the velocity of the average PomXY cluster trajectory towards midnucleoid (Fig.~\ref{fig:parameter_sweep}C). Qualitatively, large hydrolysis rates ($k_h = \SI{1}{s^{-1}}$) have essentially the same effect (Fig.~\ref{fig:parameter_sweep}C). Hence, there is a hydrolysis rate $k_h$ that minimizes the time the PomXY cluster needs to reach midnucleoid. Although the average PomXY cluster trajectory behaves similarly for large and small hydrolysis rates, we observe that the variance of the cluster distribution over time decreases with increasing hydrolysis rate (Fig.~\ref{fig:parameter_sweep}C).

Apart from the ATP hydrolysis rate, we expect the diffusivity of PomZ on the nucleoid to be a crucial parameter for cluster movement, because it determines the time available for PomZ dimers to explore the nucleoid to the left or right of the cluster. Interestingly, when we reduce the diffusivity of PomZ on the nucleoid in the simulations, the clusters begin to oscillate around the midnucleoid position (Fig.~\ref{fig:parameter_sweep}D, \ref{fig:parameter_sweep}E and \ref{S1_Fig}).
Finally, we also decreased the diffusion constant of PomZ dimers on the PomXY cluster, while keeping the diffusion constant on the nucleoid fixed. In this case, the clusters take longer to reach midcell (Fig.~\ref{fig:parameter_sweep}F). 

In addition to the parameter sweeps shown in Fig.~\ref{fig:parameter_sweep}, we also considered the PomXY cluster trajectories when the spring stiffness $k$ and the total PomZ dimer number $N_\text{total}$ are varied. In short, the cluster trajectories do not change remarkably when the spring stiffness is altered over one order of magnitude, and an increase in the particle number increases the velocity of cluster movement towards midcell (Fig.~\ref{S2_Fig}). 

To summarize, we observed that there exists an ATP hydrolysis rate that minimizes the time taken to reach midnucleoid. The diffusion constant of PomZ on the nucleoid determines whether the PomXY cluster moves towards or oscillates around midnucleoid. Moreover, the clusters move faster towards midcell if PomZ dimers diffuse faster on the PomXY cluster.

\subsection*{A deterministic approximation for the average cluster trajectory}
In order to account for the features revealed by our \textit{in-silico} parameter sweeps, we need to take a closer look at how forces are generated in the system. PomZ dimers generate forces by interacting with the PomXY cluster due to the elastic properties of the chromosome and the proteins themselves. However, fluctuations of PomZ dimers around their equilibrium positions do not produce a net force. So how is the net force that leads to the bias in cluster movement actually produced? In order to answer this question, we first note that detailed balance is broken in the model. In this non-equilibrium system, we have a cyclic flow of PomZ dimers: PomZ dimers attach to the nucleoid in their active state (as ATP-bound PomZ dimers), diffuse on the nucleoid and are released into the cytosol in their inactive state (ADP-bound PomZ monomers) at some point after encountering the PomXY cluster. This cyclic flow can lead to a net force exerted on the cluster as we describe in the following.

When the PomZ dynamics is fast compared to the PomXY cluster dynamics, as suggested by the experimental data for wild-type \textit{M. xanthus} cells \cite{Schumacher2017a}, we can make an adiabatic assumption, i.e.\ the time scales for the PomXY cluster and PomZ movements can be separated. More specifically, on the time scale of PomZ dynamics, the cluster position can be regarded as constant. Here, and in the rest of this section, we assume that the adiabatic assumption holds true, and approximate the system by a stationary model, i.e.\ a system with a fixed PomXY cluster position. 
As we neglect exclusion effects on the nucleoid, PomZ dimers can only interact with each other via the PomXY cluster. However, when the cluster is stationary, no interaction between the cluster-bound PomZ dimers is possible, and thus there are no correlations between the movements of different PomZ proteins. Therefore, we can consider the interactions of PomZ dimers with the PomXY cluster as independent, which yields the following deterministic approximation for the total net force, $F$, acting on a cluster at position $x_c$
\begin{equation}
F(x_c) = (N_R(x_c) - N_L(x_c)) f,
\label{eq:total_force}
\end{equation}
with $f$ being the time-averaged force exerted by a single PomZ dimer that attaches to the nucleoid on the right side of the cluster. For symmetry reasons, a PomZ dimer coming from the left then exerts a time-averaged force $-f$. $N_R$ and $N_L$ denote the numbers of PomZ dimers that are bound to the PomXY cluster and had originally attached to the nucleoid to the right and left of the cluster, respectively. These two numbers increase with the diffusive flux of nucleoid-bound PomZ dimers reaching the cluster region from the right and left side, $j_{R/L}$, respectively, and decrease with the ATP hydrolysis rate, $k_h$, as long as the attachment rate to the PomXY cluster is non-zero. Hence, we obtain 
\begin{equation}
N_{R/L}(x_c) = \frac{j_{R/L}(x_c)}{k_h}
\end{equation}
in the steady-state. Inserting this into Eq.~\ref{eq:total_force}, yields
\begin{align}
\label{eq:proportionality_constant}
\nonumber
F(x_c) &= \frac{j_R(x_c)-j_L(x_c)}{k_h} f \\
&= \frac{f}{k_h} j_\text{diff}(x_c) \equiv C j_\text{diff}(x_c).
\end{align}
We conclude that the net force is proportional to the flux difference of PomZ dimers at the cluster, $j_\text{diff}$, and the proportionality constant is given by $C =  f /k_h$. Importantly, simulation results with a fixed PomXY cluster position confirm the observation that the total force exerted on the PomXY cluster is proportional to the PomZ flux difference (Fig.~\ref{S3_Fig}).

Next, we investigate how the net force exerted on the PomXY cluster results in movement of the cluster. Notably, the PomZ dimers interacting with the PomXY cluster not only produce a net force on the cluster, they also reduce the mobility of the cluster by tethering it to the nucleoid. We assume that these two processes can be considered independently. When we simulate the movement of a PomXY cluster with a fixed number of PomZ dimers bound to it (the ATP hydrolysis rate $k_h$ is set to zero) and apply forces of different magnitudes to the cluster, we observe a linear increase in the steady-state velocity of the cluster with the force (Fig.~\ref{S4_Fig}). This suggests that the force exerted on the cluster is balanced by a frictional force.
The average velocity of the PomXY cluster is then determined by the flux difference of PomZ dimers into the cluster, the proportionality constant $C$ and the effective friction coefficient $\gamma(x_c)$ of the cluster
\begin{equation}
v(x_c) = \frac{F(x_c)}{\gamma(x_c)} = \frac{C j_\text{diff}(x_c)}{\gamma(x_c)},
\label{eq:vel_of_cluster}
\end{equation}
which is the central equation in our analysis. Note that the friction coefficient depends on the position of the PomXY cluster, because the number of PomZ dimers attached to the cluster changes with the cluster position. To obtain the average cluster trajectory, we need to integrate Eq.~\ref{eq:vel_of_cluster} over time. Hence, we need expressions for the flux difference into the cluster, the constant $C$ and the effective friction coefficient of the PomXY cluster.

\paragraph*{Analytical expression for the PomZ flux difference.}
To derive an analytical expression for the difference in PomZ flux into the cluster in the adiabatic limit, we introduce a reaction-diffusion (RD) model which closely resembles that investigated in the stochastic simulation. The nucleoid is modelled as a one-dimensional line of length $L$ and the PomXY cluster is a finite interval on this line, $I_c = [x_c - L_c/2, x_c + L_c/2]$. Let $c(x,t)$ denote the concentration of PomZ dimers bound to the nucleoid only, $c_b(x,t)$ the concentration of PomZ dimers bound to the nucleoid and cluster, and $N_\text{cyto}$ the number of PomZ dimers in the cytosol. The nucleoid and the PomXY cluster are assumed to have reflecting boundary conditions for the nucleoid-bound and cluster-bound PomZ dimers, respectively. In accordance with the stochastic model, PomZ dimers attach to the nucleoid left and right of the cluster with rate $k_\text{on}$ and diffuse on the nucleoid with diffusion constant $D_\text{nuc}$. 
In the RD model we simplify the interactions of PomZ dimers with the PomXY cluster: nucleoid-bound PomZ in the cluster region, $I_c$, can bind to the cluster with a rate $k_a^\text{total}$, neglecting the elasticity of the PomZ dimers and the chromosome included in our stochastic model. To obtain an expression for $k_a^\text{total}$ that resembles attachment in the stochastic model, we need to integrate the rate of attachment of PomZ dimers positioned at $x$ on the nucleoid to position $y$ on the PomXY cluster over all cluster positions: 
\begin{equation}
k_a^\text{total}(x) = \int_{x_c-L_c/2}^{x_c + L_c/2} k_a^0 e^{- \frac 1 2 \frac{k}{k_B T} (y-x)^2} dy.
\end{equation}
Since we expect the physiological value of the spring stiffness $k$ to be large (Table \ref{S1_Table}, \cite{Hu2017a}), the Boltzmann factor decays quickly, and hence we can neglect boundary effects of the PomXY cluster ($L_c \rightarrow \infty$):
\begin{equation}
k_a^\text{total} = \left\{
	\begin{array}{ll}
	k_a^0 \sqrt{\frac{2\pi k_BT}{k}}, & x \in I_c,\\
	0, & \text{otherwise}.
	\end{array}
	 \right.
\end{equation}
We assume that PomZ dimers bound to the PomXY cluster and the nucleoid diffuse with a diffusion constant $D_\text{b}$. How does this value depend on the diffusion constants of PomZ on the nucleoid and on the PomXY cluster, $D_\text{nuc}$ and $D_\text{clu}$, and on the spring stiffness, $k$, in the lattice gas model? If the diffusion constants on the cluster and the nucleoid are the same, $D_\text{clu} = D_\text{nuc} \equiv D$, the center of mass of the PomZ dimer diffuses with $D_\text{b} = 0.5 D$, independently of the spring stiffness. This result is known from the Rouse model, which models polymers as $n$ beads coupled by springs. Here, the diffusion constant of the center of mass of the polymer decreases with the number of beads as $1/n$ \cite{DeGennes1976}. For a stiff spring coupling the nucleoid and cluster binding sites, the diffusion constant of the two binding sites can be considered as equal and equivalent to that for the center of mass. 

\begin{figure*}[t]
\includegraphics[]{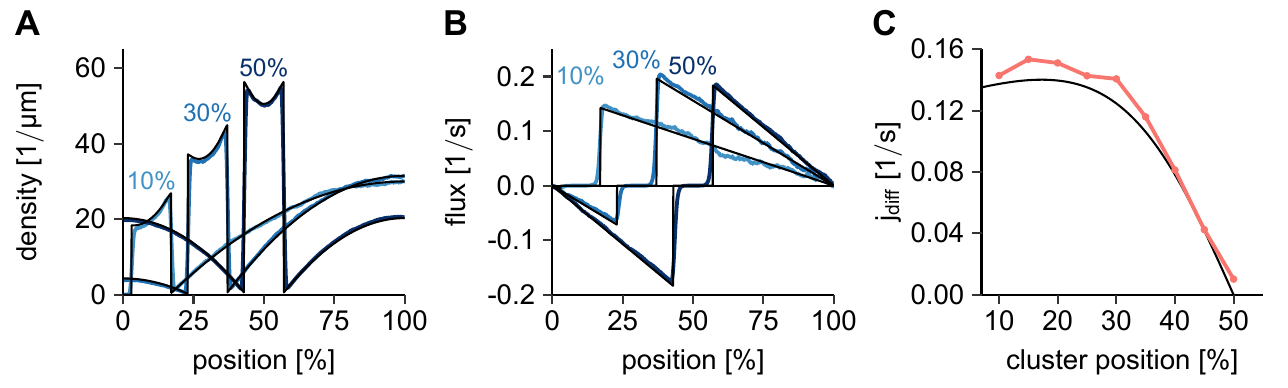}
\caption{\textbf{Comparison of the RD with the stochastic model.} (A, B) Density and flux of PomZ dimers along the nucleoid for PomXY clusters at $10\%, 30\%$ and $50\%$ nucleoid length. For the PomZ density we use the nucleoid binding site as PomZ dimer position. Regarding the flux, only PomZ dimers bound to the nucleoid, but not the PomXY cluster, are considered. The analytical result obtained from the RD equations is shown in black and the results from the stochastic simulations in blue. (C) PomZ flux difference into the cluster as a function of the cluster position. The black line indicates the result from the RD equation, the red points are results from the stochastic simulations. For the data shown in this Figure we simulated $100$ cluster trajectories with parameters as in Table \ref{S1_Table}. See the Materials and methods section for more details.}
\label{fig:comparison_rd_stochsim}
\end{figure*}

Finally, PomZ dimers bound to the cluster and nucleoid can hydrolyze ATP and subsequently detach into the cytosol with rate $k_h$. With these model assumptions we obtain the following reaction-diffusion equations to describe the PomZ dynamics, respecting particle number conservation:
\begin{alignat}{2}
\label{eq:RD_model1}
\partial_t c(x,t) &= D_\text{nuc} \partial_x^2 c(x,t) + \frac{k_\text{on} N_\text{cyto}(t)}{L}, && (x \not\in I_c)\\
\label{eq:RD_model2}
\partial_t c(x,t) &= D_\text{nuc} \partial_x^2 c(x,t) - k_a^\text{total} c(x,t), && (x \in I_c)\\
\nonumber
\partial_t c_b(x,t) &= D_\text{b} \partial_x^2 c_b(x,t) + k_a^\text{total} c(x,t) \\
\label{eq:RD_model3}
& \hspace{0.5cm}- k_h c_b(x,t), && (x \in I_c)\\
\nonumber
\partial_t N_\text{cyto}(t) &= k_h \int_{x_c-L_c/2}^{x_c+L_c/2} dx\, c_b(x,t) \\
\label{eq:RD_model4}
& \hspace{0.5cm}- k_\text{on}\frac{L-L_c}{L} N_\text{cyto}(t),
\end{alignat}
with the following no-flux boundary conditions at the nucleoid and PomXY cluster edges:
\begin{align}
\label{eq:RD_model5}
\partial_x c(x,t)|_{x=0} &= 0 = \partial_x c(x,t)|_{x=L},\\
\label{eq:RD_model6}
\partial_x c_b(x,t)|_{x=x_c-L_c/2} &= 0 = \partial_x c_b(x,t)|_{x=x_c+L_c/2}.
\end{align}
We solved the stationary state of the reaction-diffusion system analytically using \textit{Mathematica} \cite{Mathematica} (for details see SI Text \ref{S1_File}). The results obtained from the RD equations for the PomZ density and flux on the nucleoid, as well as the flux difference into the cluster, agree with the stochastic simulation results for the parameter values given in Table \ref{S1_Table} (Fig.~\ref{fig:comparison_rd_stochsim}).
A large spring stiffness is necessary for the good agreement between the two different models, because in the RD model we do not include the elasticity of the PomZ dimers, which is best reflected by a large spring stiffness. We observe that the density and the flux profiles are asymmetric for off-center clusters, and become more and more symmetric as the cluster approaches midnucleoid (Fig.~\ref{fig:comparison_rd_stochsim}A and \ref{fig:comparison_rd_stochsim}B). This leads to a diffusive flux difference into the cluster that decreases towards midnucleoid (Fig.~\ref{fig:comparison_rd_stochsim}C). The slight deviation between the flux difference from the stochastic and the RD model is due to the fact that in the simulations the PomZ dimers react rapidly, but not instantaneously (as assumed in the adiabatic limit) to a change in the PomXY cluster position. If the PomXY clusters are kept fixed in the simulations, the simulated flux difference matches perfectly with the results from the RD equations (Fig.~\ref{S3_Fig}). Note that the flux difference decreases slightly for clusters close to the nucleoid pole, which is due to total particle number conservation in the system. 

\paragraph*{Force exerted by a single PomZ dimer.}
Next, we investigate the force exerted by a single PomZ dimer on the PomXY cluster. How can the interaction of a PomZ dimer with the PomXY cluster lead to a net force? First, PomZ dimers can exert a net force by attaching to the PomXY cluster in a stretched configuration: a particle to the left/right of the cluster can bind to the cluster from a position beyond either end (Fig.~\ref{fig:stochastic_model}B). Second, PomZ dimers interacting with the PomXY cluster can diffuse on both the nucleoid and cluster. When they reach the edge of the PomXY cluster they can impart a force to the cluster, because the cluster binding site is restricted in its movement, in contrast to the nucleoid binding site (Fig.~\ref{fig:stochastic_model}B). 
In principle, PomZ dimers can also generate a net force at the nucleoid ends, which moves the PomXY cluster towards midnucleoid, but since PomZ dimers rarely encounter the nucleoid ends, this can only represent a minor contribution. To investigate the force generated by a single PomZ dimer, we performed stochastic simulations with a stationary PomXY cluster and only one PomZ dimer in the system, which can attach to the nucleoid at a site so far away from the cluster that no interaction between the particle and the cluster is possible. Once bound to the nucleoid, the PomZ dimer can diffuse along it, interact with the PomXY cluster and detach. 
We record the positions of the PomZ dimer binding site on the nucleoid and on the PomXY cluster when the particle attaches to the cluster. Furthermore, we record the time-averaged and time-integrated force exerted by a PomZ dimer on the PomXY cluster. Note that the position of the cluster relative to the nucleoid is irrelevant as long as the cluster is not positioned close to a nucleoid boundary. We only consider particles that attach to the nucleoid at the right side of the cluster. For PomZ dimers that attach to the left side only the sign of the forces changes due to a left-right symmetry of the system. 

Importantly, the ensemble average of the time-integrated forces of PomZ dimers attaching from the right is greater than zero, $f_\text{int} = \SI[separate-uncertainty]{0.1338(00009)}{\pico\newton\s}$ (the error denotes the standard error of the mean), showing that a particle coming from the right indeed, on average, exerts a force directed to the right and vice versa (Fig.~\ref{S5_Fig}, parameters as in Table \ref{S1_Table}). The average distance between nucleoid and cluster binding sites when the particle attaches to the PomXY cluster is $\Delta x_0 = \SI{0.0034}{\um}$, which is less than one lattice spacing, $a = \SI{0.01}{\um}$. With this initial distance between the two binding sites, we can estimate how much force is produced by the binding of a PomZ dimer to the PomXY cluster in a stretched configuration. Let us assume that the cluster binding site is fixed. Then the movement of the nucleoid binding site can be considered as an Ornstein-Uhlenbeck process, i.e.\ a particle diffusing in a potential given by the energy of the spring. This leads to the following estimate for the time-integrated force
\begin{align}
\nonumber
f^\text{OU}_\text{int} &= \int_0^{1/k_h} k \Delta x_0 e^{-\beta k D_\text{nuc}t} dt \\*
&\approx \frac{\Delta x_0}{\beta D_\text{nuc}} \approx \SI{1.36 E-3}{\pico\N\s},
\end{align}
which is two orders of magnitudes smaller than the time-integrated force obtained from the simulations. Note that the time scale for relaxation in the Ornstein-Uhlenbeck process $\tau_\text{OU} = 1/(\beta k D_\text{nuc}) = \SI{0.01}{s}$ is much less than the time for which a PomZ dimer is typically attached to the PomXY cluster $\tau = 1/k_h = \SI{100}{s}$ for the parameter set in Table \ref{S1_Table}. Hence, we conclude that the force produced due to the initial deflection of the PomZ dimer is not the main contribution to the time-integrated force for the parameters considered here. This suggests that the force exerted by a PomZ dimer when it encounters the PomXY cluster's edge is an important contribution to the net force generated. 

The constant $C$ in Eq.~\ref{eq:vel_of_cluster} is given by the time-averaged one-particle force $f$ divided by the ATP hydrolysis rate $k_h$. In our simulation, we determined the ensemble average of the time-averaged forces using the interaction times as weights, which results in a constant $C = \SI{0.1335}{\pico\newton\s}$ for the parameters as in Table \ref{S1_Table} (see Materials and methods for details). Note that this value for $C$ is very similar to the proportionality constant between the total force exerted on the PomXY cluster and the PomZ flux difference for a stationary cluster, $C = \SI{0.1348}{\pico\newton\s}$ (Fig.~\ref{S3_Fig}).
We now have approximations for the flux difference into the cluster (results from Eq.~\ref{eq:RD_model1}-\ref{eq:RD_model6}) and the proportionality constant between the force and the flux difference (Eq.~\ref{eq:proportionality_constant}). The only parameter yet to be estimated is the effective friction coefficient of the PomXY cluster, which we consider next.

\begin{figure*}[t]
\includegraphics[]{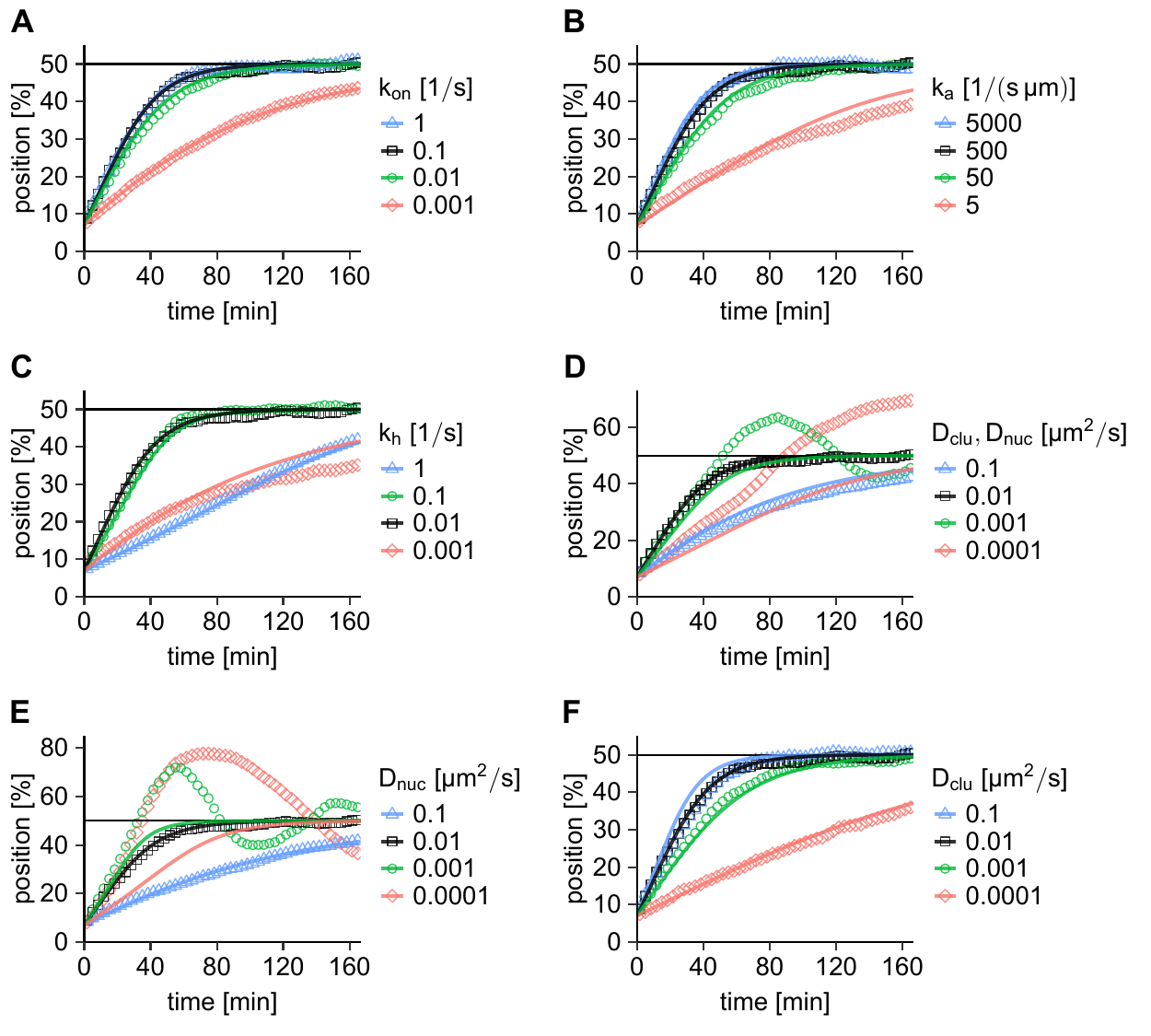}
\caption{\textbf{Comparison of the average cluster trajectory from simulations with our semi-analytical approximation.} (A-F) The cluster trajectories obtained from integrating the equation of motion of the PomXY cluster, Eq.~\ref{eq:vel_of_cluster}, (solid lines) agree with the simulation results for most parameters (points of different shape, same data as shown in Fig.~\ref{fig:parameter_sweep}). In the semi-analytical approximation we use the theoretical values for the flux difference and the friction coefficient together with the simulated value for $C$. For the parameters for which the cluster overshoots midcell (small $D_\text{nuc}$), our semi-analytical theory does not match the simulation results. This is expected, because we make the assumption that the PomZ dimer dynamics is faster than PomXY cluster movement (adiabatic assumption). If not explicitly given in the Figure, the parameters are as in Table \ref{S1_Table}.}
\label{fig:average_cluster_trajectory}
\end{figure*}

\paragraph*{Effective friction coefficient of the PomXY cluster.}
We derived an analytical expression for the effective friction coefficient by assuming that the PomXY cluster and the nucleoid boundaries can be disregarded (see SI Text \ref{S2_File}). We find that the effective friction coefficient of the PomXY cluster is given by the cytosolic friction coefficient plus a term that increases linearly with the number, $N$, of PomZ dimers bound to the cluster:
\begin{equation}
\label{eq:effective_friction}
\gamma (x_c) = \gamma_c + \frac{k_BT N(x_c)}{D_\text{clu} + D_\text{nuc}}.
\end{equation}
Our analytical result agrees with the simulation results for an infinitely extended cluster and nucleoid and a constant number $N$ of PomZ dimers bound to the cluster (Fig.~\ref{S6_Fig}, for details see Materials and methods section). In general, an approximation for the number of cluster-bound PomZ dimers can be obtained from the stationary solution of the RD model (see SI Text \ref{S1_File}). With the friction coefficient of the PomXY cluster we now have estimates for all factors that contribute to the velocity of the PomXY cluster (Eq.~\ref{eq:vel_of_cluster}) and hence determine the average cluster trajectory. 

\paragraph*{Semi-analytical approach explains observed simulation results.}
Using the analytical values for the PomZ flux difference at the cluster boundaries, the effective friction coefficient of the PomXY cluster, and the simulated values for the force exerted by a single particle on the PomXY cluster, we can obtain an estimate for the average cluster velocity. The single particle force, and thus the constant $C$, does not change with the cluster position. Hence the dependence of the velocity on the position of the cluster is given by an analytical expression, which can be integrated numerically. 
For most of the parameters, the simulated average cluster trajectory and the approximation from our semi-analytical approach are in good agreement (Fig.~\ref{fig:average_cluster_trajectory}A-\ref{fig:average_cluster_trajectory}F, \ref{S7_Fig}). In some cases, e.g.\ for $k_h = \SI{0.001}{\per\second}$, our approximation lies above the simulation results. This is probably due to the fact that the dynamics of the PomXY cluster and the PomZ dimers cannot really be separated, as assumed in our approximation. If PomZ dimers remain attached to the PomXY cluster for a long time, which is the case for low ATP hydrolysis rates, the cluster can move a certain distance before the PomZ dimers hydrolyze ATP and detach from the cluster and the nucleoid. If the cluster is to the left of midnucleoid and moves to the right, the cluster-bound PomZ dimers move on average to the left and accumulate at the left boundary of the cluster. This leads to an increase in the force exerted by the PomZ dimers on the left edge of the cluster, and hence reduces the velocity of the cluster's movement towards midnucleoid. 

\begin{figure*}[t]
\includegraphics[]{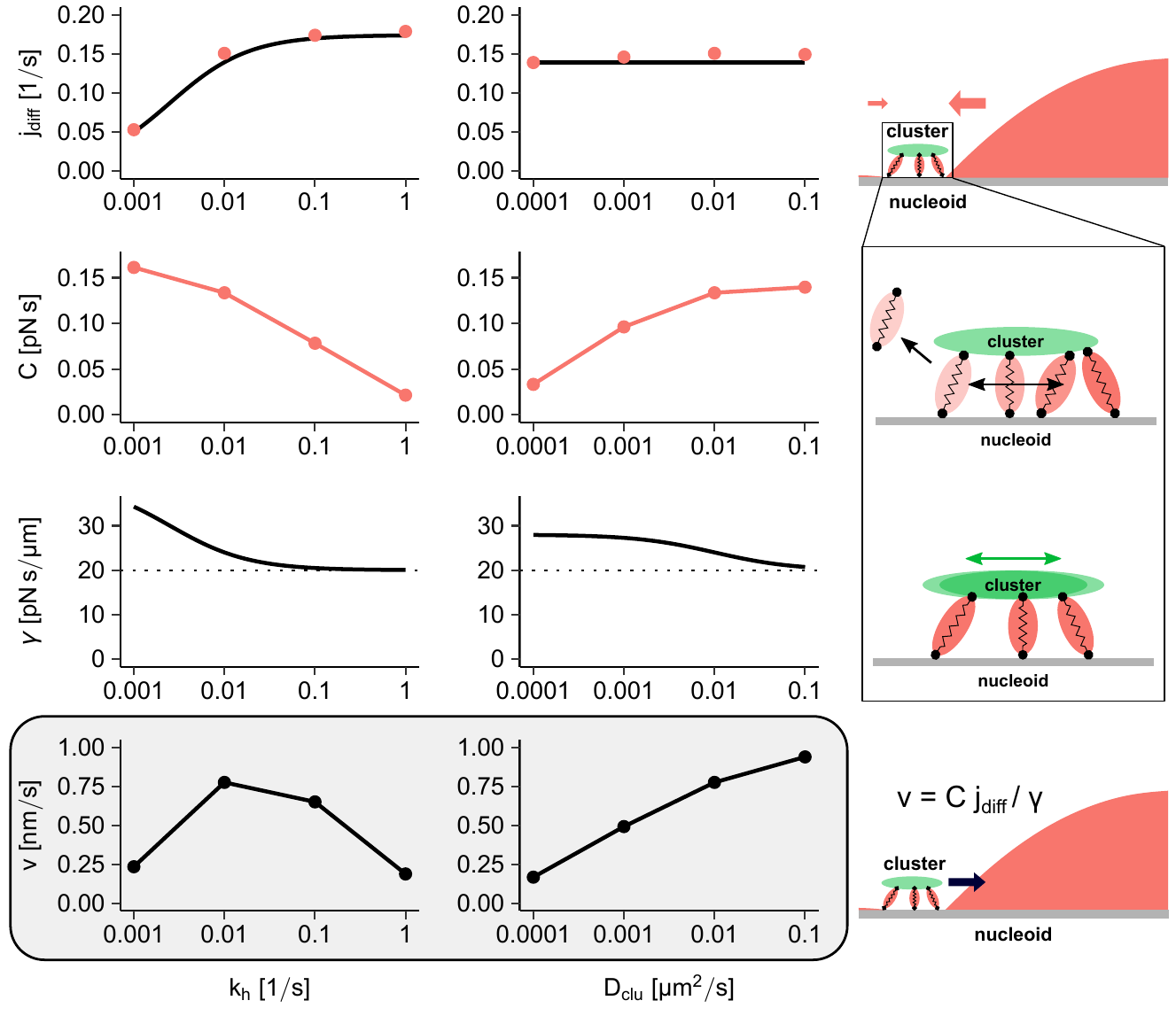}
\caption{\textbf{Force generation in the flux-based model.}
The average velocity of the cluster is approximated by the difference in flux of PomZ dimers into the cluster region from either side, $j_\text{diff}$, the constant $C$, which describes the force exerted by a single PomZ dimer on the PomXY cluster, and the effective friction coefficient of the PomXY cluster, $\gamma$. Here, the impact of varying the hydrolysis rate $k_h$ (first column) or the diffusion constant of PomZ dimers on the PomXY cluster, $D_\text{clu}$ (second column) is shown. The first row shows the PomZ flux difference at the cluster when the cluster is at $20\%$ nucleoid length. The result from the RD equations (black line) matches the stochastic simulation results (red points). 
The second row shows the proportionality constant $C$ determined from one-particle simulations (more than \num{30000} PomZ dimer-cluster interactions are simulated). The points are connected by lines to guide the eye. The third row shows the analytical curves for the effective friction coefficient of the PomXY cluster at $20\%$ nucleoid length obtained from Eq.~\ref{eq:effective_friction}. An increase in the number of PomZ dimers bound to the PomXY cluster (e.g.\ for low $k_h$ values) or increased friction of PomZ dimers on the PomXY cluster (e.g.\ for low $D_\text{clu}$ values) leads to effective friction coefficients larger than the cytosolic friction coefficient (dotted horizontal line). Finally, the average velocity of the cluster can be calculated based on the flux difference, the constant $C$ and the friction coefficient using Eq.~\ref{eq:vel_of_cluster}. The velocity obtained using the theoretical values for both the flux difference and the friction coefficient, and the simulated values for $C$, is shown in the last row (grey box). The points are connected by lines to guide the eye. If not explicitly given in the Figure, the parameters are as in Table \ref{S1_Table}. See the Materials and methods section for more details.}
\label{fig:force_generation}
\end{figure*}

When the PomZ dynamics is slow compared to the PomXY cluster dynamics, as is the case for small $D_\text{nuc}$, the adiabatic assumption breaks down, and our semi-analytical approach fails to reproduce the simulated cluster trajectories (Fig.~\ref{fig:average_cluster_trajectory}D and \ref{fig:average_cluster_trajectory}E). With the semi-analytical approximation, we dissected the net force generation into its parts, which allows for a more detailed analysis of the cluster dynamics. The \textit{in-silico} parameter sweeps revealed an interesting behavior of the cluster trajectories when the ATP hydrolysis rate and the diffusion constant of PomZ on the PomXY cluster are varied (Fig.~\ref{fig:parameter_sweep}C and \ref{fig:parameter_sweep}F). In the first case, there is an optimal hydrolysis rate that minimizes the time required for the cluster to reach midnucleoid. In the second case, movement of the cluster becomes less directed towards midnucleoid when the mobility on the PomXY cluster is decreased. Next, we will investigate the cluster dynamics in these two examples by considering the different contributions to the cluster's velocity separately. 

Fig.~\ref{fig:force_generation} gives an overview of the different contributions to the cluster's velocity when the ATP hydrolysis rate, $k_h$, or the diffusion constant of PomZ on the PomXY cluster, $D_\text{clu}$, is varied (for further parameters see Fig.~\ref{S8_Fig} and \ref{S9_Fig}). The flux difference of PomZ dimers into the cluster ``measures'' the position of the cluster on the nucleoid (first row). Interactions of PomZ dimers with the PomXY cluster lead to forces that are exerted on the cluster. Cluster-bound PomZ dimers exert a net force upon encountering the PomXY cluster's edge and they increase the friction of the PomXY cluster by tethering it to the nucleoid (second and third row). Taken together, a difference in the PomZ fluxes onto the ends of the cluster and local force generation by PomZ dimers at the PomXY cluster boundaries impart a velocity to the cluster that leads to a net movement towards midnucleoid (fourth row). 

Increasing the hydrolysis rate increases the flux difference, because the ATP hydrolysis rate determines the rate of PomZ dimer release from the nucleoid, and hence is important for the flux of PomZ dimers through the system (Fig.~\ref{fig:force_generation}). To understand the dependence of $C = f/k_h$ on $k_h$, we first consider the dependence of the time-averaged force $f$ on $k_h$. When a PomZ dimer attaches to the PomXY cluster, it typically binds close to the PomXY cluster's edge. The probability density of the particle flattens over time because the PomZ dimer diffuses on the PomXY cluster and the nucleoid. For very long times, the average PomZ dimer position is the center of the PomXY cluster. Hence a particle is more likely to impart a net force to the PomXY cluster early in the interaction period than late. We conclude that the ensemble average of the time-averaged forces of a single PomZ dimer increases with $k_h$, which we indeed observe in the simulations (Fig.~\ref{S10_Fig}). 
Nevertheless, this increase is less than linear in $k_h$, such that $C = f/k_h$ decreases with $k_h$ (Fig.~\ref{fig:force_generation}). The effective friction coefficient also decreases with increasing $k_h$, because the number of PomZ dimers bound to the PomXY cluster decreases. 
The increasing flux difference, the decreasing constant $C$, and a decrease in friction together result in a maximal velocity, $v$, of the cluster for intermediate $k_h$ values (Fig.~\ref{fig:force_generation}). This explains why there exists a hydrolysis rate for which the cluster trajectory reaches midnucleoid in a minimal time. Furthermore, we observed that the variance in the cluster position decreases with increasing $k_h$. Since an increase in the hydrolysis rate increases the flux of PomZ dimers through the system and decreases the interaction time of PomZ dimers with the PomXY cluster, we expect a less stochastic movement of the cluster for larger hydrolysis rates, as observed. 

Furthermore, we considered the case where the PomZ diffusion constant on the PomXY cluster is reduced while keeping the diffusion constant on the nucleoid fixed. Since diffusion on the PomXY cluster only affects the PomZ dynamics locally at the PomXY cluster, changing this rate does not change the flux difference of PomZ into the cluster (Fig.~\ref{fig:force_generation}), but it does alter the magnitude of force generation at the cluster. We find that the time-averaged one-particle force decreases with decreasing diffusion constant (Fig.~\ref{fig:force_generation}), which explains the increase in the time required for a cluster to reach midnucleoid for small diffusion constants (Fig.~\ref{fig:parameter_sweep}F). Why the force decreases when the diffusion constant on the PomXY cluster is reduced can be understood intuitively as follows: Our findings indicate that the main contribution to the net force generated by the PomZ dimers is the force they exert when they encounter the PomXY cluster's edge. When the diffusion constant of PomZ on the PomXY cluster, $D_\text{clu}$, is zero, the nucleoid binding site of a cluster-bound PomZ dimer equilibrates and fluctuates around this equilibrium position without producing a net force. Hence, in this case only the attachment of PomZ dimers to the PomXY cluster in a stretched state results in a net force. Note that for a non-zero diffusion constant on the PomXY cluster two opposing effects play a role for the number of times the PomZ dimers reach the PomXY cluster's edge: The smaller $D_\text{clu}$, the higher the probability of finding the cluster binding site close to the cluster's edge, but the lower the chance that the cluster binding site will hop to the cluster's edge, or indeed will hop at all. 

To summarize, with our semi-analytical approach we can separate the global asymmetry, i.e.\ a cluster located at an off-center position, which results in different diffusive PomZ fluxes into the cluster, from the forces locally exerted on the cluster. In particular, we can identify the different contributions to the velocity of the cluster and thereby understand why there is an ATP hydrolysis rate that results in a minimal time the clusters need to reach midcell and why diffusion of PomZ dimers on the PomXY cluster matters in our model.

\subsection*{Oscillatory behavior vs.\ midnucleoid localization of the cluster}

We observe a marked discrepancy between the simulated average cluster trajectory and our approximation when the diffusion constant of PomZ on the nucleoid is reduced and the cluster oscillates around midnucleoid (Fig.~\ref{fig:average_cluster_trajectory}D and \ref{fig:average_cluster_trajectory}E). Deviations from our theoretical predictions are to be expected in this situation, because we make an adiabatic assumption in our semi-analytical approach, i.e.\ we assume that the PomZ dimer dynamics on the nucleoid is fast compared to the cluster movement. This assumption no longer holds when PomZ dimers diffuse slowly on the nucleoid. The distribution of PomZ density along the nucleoid determined from simulations with a dynamic cluster deviates drastically from its steady-state distribution (Fig.~\ref{S11_Fig}). If the cluster initially lies to the left of midnucleoid and approaches midnucleoid from that side, our theory predicts a symmetric PomZ density, whereas the simulations show a higher density in front of the cluster and also an asymmetric density of PomZ dimers at the PomXY cluster. The flux difference also deviates from the stationary case: it increases as the cluster moves towards midnucleoid instead of vanishing at midnucleoid (Fig.~\ref{S11_Fig}). Both the asymmetric density and the non-zero flux difference at midnucleoid are in accordance with the observed oscillatory behavior. 

\begin{figure*}[t!]
\includegraphics[]{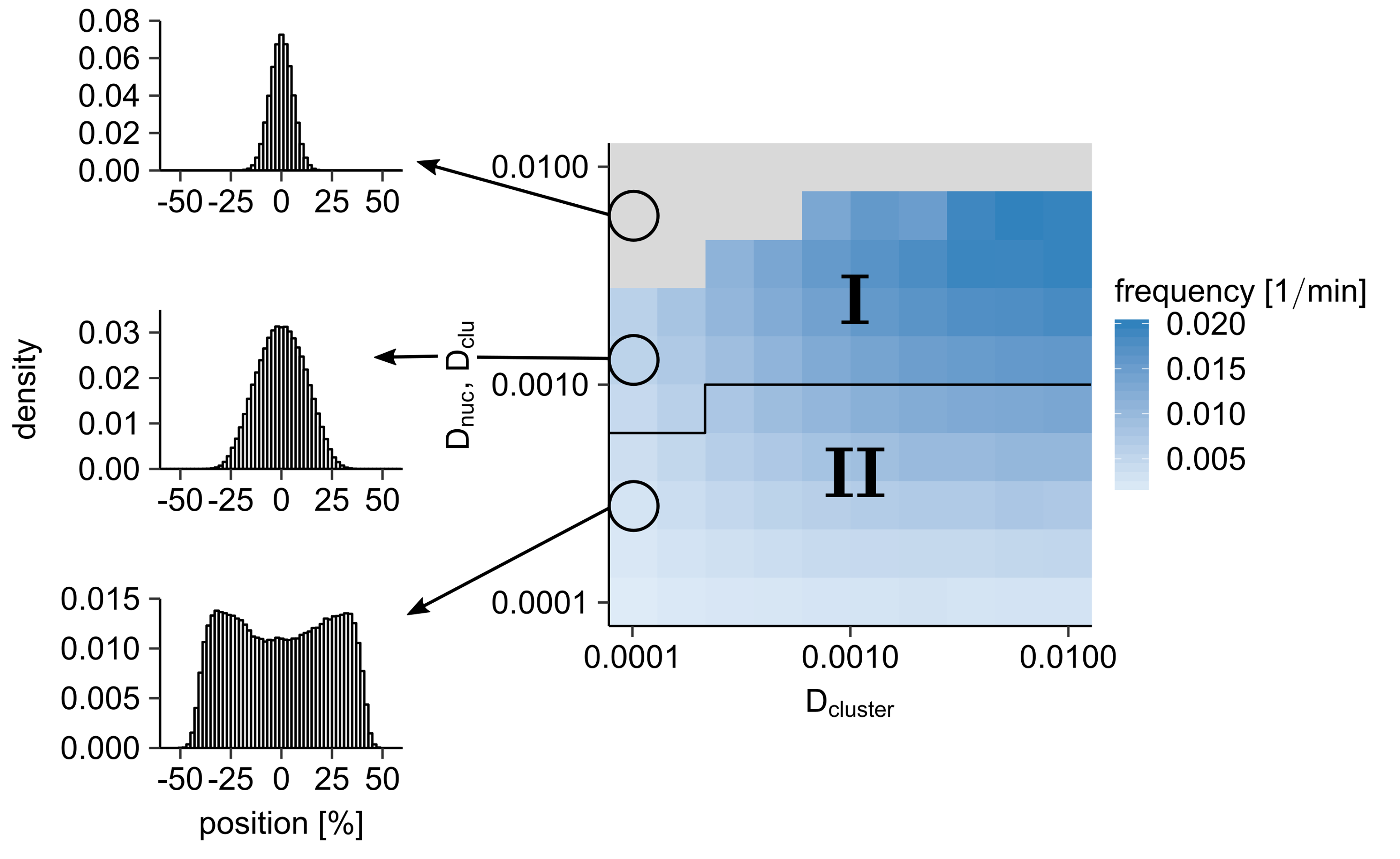}
\caption{\textbf{Oscillatory cluster movement occurs if PomZ dynamics is slower than PomXY cluster dynamics.} We varied the diffusion constant of the PomXY cluster, $D_\text{cluster} = k_BT/\gamma_c$, and the diffusion constant of PomZ on the nucleoid and PomXY cluster ($D_\text{clu}$ and $D_\text{nuc}$, are set to the same value). The other parameters are as in Table \ref{S1_Table}. The clusters localize at midnucleoid for high PomZ diffusion constants and low diffusion constants of the PomXY cluster $D_\text{cluster}$ (grey region). If the diffusion constant of PomZ is decreased from $\SI{0.01}{\um^2/s}$, the clusters begin to oscillate, because the time scales of the PomZ dimer dynamics and the PomXY cluster dynamics become comparable (region I). The average frequency of oscillation is shown in blue (100 runs per parameter set are considered). In this parameter regime, the distribution of cluster positions is peaked around midnucleoid (see histograms on the left for the parameters marked with circles). For even lower PomZ diffusion constants (region II) the cluster positions are bimodally distributed. In the simulations, the clusters begin at midnucleoid and are recorded for \SI{10000}{\minute}. The black line in the frequency plot indicates a threshold. Below the curve the cluster distribution is bimodal, above it the distribution has only one peak. For details of the data analysis see the Materials and methods section. }
\label{fig:oscillations}
\end{figure*}

The switch between cluster localization at midnucleoid and oscillatory movement around midnucleoid is regulated by the relative time scales of PomZ dynamics and cluster dynamics: If the PomXY cluster is moving slowly or the PomZ dimers move fast, the latter have time to adjust to a change in the PomXY cluster position. On the other hand, if the PomXY cluster moves fast or the PomZ dimers move slowly, the PomZ dimer distribution deviates from the stationary case. The delay between the movement of the PomXY cluster and the build-up of the PomZ gradient, which in turn biases the movement of the cluster, leads to oscillations: the longer the delay, the larger the amplitudes of the oscillations. To investigate the oscillatory case further, we performed additional simulations in which the diffusion constant of the PomZ dimers and that of the PomXY cluster in the cytosol - which is inversely proportional to the friction coefficient, $\gamma_c$, according to the Stokes-Einstein relation - were varied. As expected, we find oscillatory behavior of the clusters for low diffusion constants of PomZ on the nucleoid and PomXY cluster
(Fig.~\ref{fig:oscillations}). In the oscillatory regime we find both bimodal and monomodal cluster position distributions (Fig.~\ref{fig:oscillations}). 

As mentioned above, the onset of oscillations depends on the time scales of PomZ gradient formation and cluster movement. In order to understand how the parameters change the behavior of the cluster trajectory, i.e.\ lead to oscillatory movement or midcell positioning, we assume that the cluster is located at midnucleoid and search for a stability condition that distinguishes the two behaviors. The diffusion time for a PomZ dimer to explore a nucleoid of size $L$ is given by
\begin{equation}
t_\text{PomZ} = \frac{L^2}{D_\text{nuc}}.
\end{equation}
In theory, the velocity of a cluster that starts from midnucleoid should be zero, because there should be no difference between the fluxes of PomZ dimers from both sides. However, due to stochastic effects, more particles may attach to the cluster from the right than from the left side, which will displace the cluster to the right. For our time scale argument, we consider an extreme case: we assume that PomZ dimers only arrive from one side, which we choose to be the right side without loss of generality. The time required for a cluster to move the length of the nucleoid is then given by
\begin{equation}
t_\text{cluster} = \frac L v \approx \frac{L \gamma(0.5 L)}{C j_\text{right}(0.5 L)},
\label{eq:tcluster}
\end{equation}
with $j_\text{right}$ being the flux of PomZ dimers into the cluster from the right. Here, we approximate the velocity of the cluster by its effective description, Eq.~\ref{eq:vel_of_cluster}, using $x_c = 0.5L$, and replace the flux difference with the flux from the right only. 
According to Eq.~\ref{eq:tcluster}, the condition for stable positioning of the cluster at midnucleoid
\begin{equation}
t_\text{PomZ} \ll t_\text{cluster} 
\end{equation}
yields
\begin{align}
\nonumber
\frac{D_\text{nuc}}{L} &\gg \frac{C j_\text{right}(0.5 L)}{\gamma(0.5 L)} \\
&= \frac{C j_\text{right}(0.5 L)}{\gamma_c + (k_BT N(0.5L))/(D_\text{clu} + D_\text{nuc})}.
\label{eq:stability_condition}
\end{align}
For the parameter sweeps considered before (Fig.~\ref{fig:parameter_sweep} and \ref{S2_Fig}), we find $t_\text{cluster} \gg t_\text{PomZ}$ for all cases except the oscillatory ones. 

With our time-scale argument, Eq.~\ref{eq:stability_condition}, we can make further predictions as to which parameters should result in oscillations. First, we consider a change in the total particle number, $N_\text{total}$. Both $j_\text{right}$ as well as the number of cluster-bound proteins, $N$, are proportional to $N_\text{total}$, and $C$ does not depend on $N_\text{total}$. Therefore, the right-hand side of Eq.~\ref{eq:stability_condition} is proportional to $N_\text{total}$ for small values of $N_\text{total}$ and converges to a constant for large values. From this we expect that oscillatory behavior may occur for large particle numbers. Simulations with $500$ PomZ dimers and a higher hydrolysis rate compared to the parameters in Table \ref{S1_Table} ($k_h = \SI{0.1}{\s^{-1}}$) indeed show oscillatory behavior, whereas simulations with the same parameters, but $100$ PomZ dimers show midnucleoid localization (Fig.~\ref{S12_Fig}). However, for very large PomZ dimer numbers we expect exclusion effects, which are not considered here, to have an impact that will also affect the cluster dynamics.

Second, we investigate the effects on cluster dynamics of changing the nucleoid length $L$. Again, the constant $C$, which represents the force exerted by a single PomZ dimer on the PomXY cluster, does not depend on $L$. The number of cluster-bound proteins decreases with $L$, because the total PomZ dimer number in the system is constant. The flux $j_\text{right}$ also decreases with $L$, because on longer nucleoids the PomZ dimers must diffuse, on average, a longer distance from their initial attachment point on the nucleoid until they reach the cluster. Bringing all terms in Eq.~\ref{eq:stability_condition} that depend on $L$ to the right hand side yields a curve that first increases with $L$, then reaches a maximum and decreases again for large $L$. Hence, we expect no oscillations for small and large nucleoid lengths and oscillations for intermediate lengths. 
Simulations with intermediate and large nucleoid lengths $L$ indeed show this behavior (Fig.~\ref{S12_Fig}).

\section*{Discussion}
We analyzed how the cluster movement changes when the rates for the key biological processes are varied over a broad range.
We found that there exists an optimal ATP hydrolysis rate of PomZ such that the time the cluster needs to move to midnucleoid is minimized. A parameter sweep of the diffusion constant of PomZ on the PomXY cluster shows that the mobility of PomZ dimers on the PomXY cluster is important for cluster movement towards midnucleoid. Qualitative changes in the cluster trajectories are observed when the diffusion constant of PomZ on the nucleoid is reduced: midnucleoid positioning of the cluster switches to oscillatory behavior of the cluster around midnucleoid. Hence, we conclude that positioning of the cluster in the flux-based model critically depends on the time scale for the cluster dynamics in comparison to the one for the PomZ dimer dynamics on the nucleoid. If the latter is slow compared to the PomXY cluster dynamics, the cluster will oscillate around midnucleoid. In contrast, fast PomZ dynamics on the nucleoid leads to midnucleoid localization of the cluster. In the latter case, the average velocity of the PomXY cluster can be described by the PomZ flux difference into the cluster, which measures how far away the cluster is from midnucleoid, the force exerted by a single PomZ dimer on the cluster, and the effective friction coefficient of the cluster, which depends on the number of PomZ dimers bound to it. 

Our mathematical model for midnucleoid localization in \textit{M. xanthus} incorporates the finding of an asymmetric protein (ATPase) flux profile on the nucleoid for a cargo positioned off-center, introduced by Ietswaart et al. \cite{Ietswaart2014}, and also the elasticity of the nucleoid as in the DNA-relay model \cite{Lim2014}. However, in \textit{M. xanthus} cells there is no clear depletion zone in the wake of the cluster and the PomZ concentration is highest at the cluster \cite{Schumacher2017a}, which is at variance with observations for ParAB\textit{S} systems, and has to be accounted for in our model. The reason why no clear depletion zone exists is probably that PomZ dimers diffuse quickly on the nucleoid \cite{Schumacher2017a}.
To account for the observation of a high PomZ dimer density at the PomXY cluster, the ATP hydrolysis rate has to be small, which is in accordance with experimental observations \cite{Schumacher2017a}, and we assume in our model that cluster-bound PomZ dimers can diffuse on both the nucleoid and the PomXY cluster. In this way, PomZ dimers are captured at the PomXY cluster and exert a force on the PomXY cluster not only due to attachment in a stretched configuration, but also when they encounter the cluster's edge while they are bound. The high PomZ density at the PomXY cluster also increases the friction of the PomXY cluster, due to the PomZ dimer interactions. That intracellular positioning and oscillations of a cargo can be produced by the same mechanism simply by varying specific system parameters is consistent with findings by Surovtsev et al. \cite{Surovtsev2016a} and Walter et al. \cite{Walter2017} for ParAB\textit{S} systems.

The flux-based model presented in this paper reproduces midnucleoid localization of a cargo and predicts oscillations around midnucleoid if PomZ dimer dynamics is slower than cluster dynamics. Apart from the spring stiffness, which is an effective constant to account for the elasticity of proteins and the chromosome, all model parameters relate to a biological process, which makes the results from parameter sweeps easy to interpret. Note, however, that the model presented here is one-dimensional, i.e.\ we do not include geometric effects due to the three-dimensional nature of the cell, the nucleoid and the Pom cluster. With our semi-analytical approach we gain insights into net force generation in a stochastic, non-equilibrium system. 

Our flux-based model makes several predictions, which would be interesting to test experimentally: One key model prediction is that the clusters start to oscillate if PomZ dimers diffuse slowly on the nucleoid. 
Furthermore, we predict that there is an optimal ATP hydrolysis rate to minimize the time the cluster takes to reach midnucleoid. Decreasing the rate of ATP hydrolysis by PomZ dimers associated with the PomXY cluster in experiments reduces the velocity of cluster movement towards midcell \cite{Schumacher2017a}. 
It would be interesting to test whether the velocity of the cluster is also reduced for an enhanced ATP hydrolysis rate in \textit{in-vivo} experiments. Further experimental insights regarding the molecular force generation mechanism in the Pom system is needed to refine the modelling in this regard. Moreover, measurements of the biological rates, such as the nucleoid attachment rate, the diffusion constant, the cluster binding rate and the ATP hydrolysis rate of ATP-bound PomZ dimers \textit{in vivo}, would help to convert the model into a quantitative one.

The research presented here gives insights into intracellular positioning mechanisms and, more generally, into transport mechanisms that are based on fluxes of particles. It explains how different cargo dynamics can be obtained by changing the time scales for the ATPase (here PomZ) dynamics and the cargo dynamics. The approach used here to dissect the net force generation in the system might prove to be useful also for ParAB\textit{S} systems. 

\section*{Materials and methods}
The mathematical model is implemented using a Gillespie algorithm \cite{Gillespie2007}, a stochastic simulation algorithm. In short, this algorithm works as follows: In each simulation step, all possible actions with their corresponding rates are determined. If the rates are constant in time, the time until any of these actions happens is exponentially distributed with the sum of all rates as rate parameter. To perform one simulation step, a uniformly distributed random number $\xi \in (0,1]$ is drawn, which results in a time step 
\begin{equation}
\Delta t = -\frac{\ln \xi}{\alpha},
\label{eq:Gillespie_time_step}
\end{equation}
where $\alpha$ is the sum over all rates. Then a uniformly distributed random number is drawn to determine which of the possible actions happens. This is done by weighting the different actions according to their rates.

Two different kinds of simulations are performed: In the first, the PomZ dynamics and the PomXY cluster dynamics are simulated (``dynamic cluster simulations''). In the second, the PomXY cluster position is kept fixed and only the PomZ dynamics is considered (``stationary cluster simulations''). In the latter case, all rates in the model are constant and the time step for the Gillespie algorithm can be calculated as described above, Eq.~\ref{eq:Gillespie_time_step}. In contrast, if the PomXY cluster is dynamic, the rates for attachment of a nucleoid-bound PomZ dimer to the PomXY cluster and the hopping rates on the nucleoid, or cluster for cluster-bound PomZ dimers, depend on the cluster position, which is itself time-dependent. The time that elapses before the next action is now given by 
\begin{equation}
\int_t^{t + \Delta t} \alpha(t') dt' = -\ln(\xi),
\end{equation}
which must be solved for $\Delta t$. Since an analytical integration of the time-dependent rates is not feasible, the expression needs to be solved numerically, which is computationally costly. However, if the PomXY cluster moves only a small distance between two Gillespie steps, the time-dependent rates also change only slightly. We tested the importance of the time dependence of the rates by approximating the time-dependent rates with their rate at time $t$, and added a rate to the simulation that has no effect, except that the time step preceding the next action is decreased on average. The results obtained when this rate was set to a high value were very similar to those found in its absence. Hence for the parameters we consider in this work, the time dependence of the rates can be ignored. 

\subsection*{Dynamic cluster simulations}
In the simulations to determine the cluster dynamics, all PomZ dimers are initially in the cytosol. The PomXY cluster position is kept fixed for $t_\text{min} = \SI{10}{\min}$ with all possible actions of the PomZ dimers allowed. As a result, the initial condition resembles the stationary distribution of PomZ dimers. The initial position of the cluster is such that the left edge of the cluster and the nucleoid coincide.

To derive PomZ flux and density profiles at specific PomXY cluster positions, the simulated fluxes and densities are recorded only if the PomXY cluster is within a certain distance of a predefined position of interest. For example, to get the PomZ flux / density for clusters at $20\%$ nucleoid length, recording begins when the PomXY cluster is in the region $20 \pm 0.2 \%$ and stops if it leaves the region $20 \pm 1\%$. Averaging is performed over all times at which the PomXY cluster resides within the specific region, weighting each density or flux profile with the corresponding time spent by the PomXY cluster in that specific region. To estimate the difference in PomZ flux into the PomXY cluster from either side, the maximal and minimal flux values in the average flux profile of PomZ dimers bound to the nucleoid, but not the PomXY cluster, are determined. These values are typically found a short distance from the edge of the PomXY cluster region, because PomZ dimers can attach to the PomXY cluster in a stretched configuration. The two extreme flux values of opposite sign are added together to get the average flux difference of PomZ dimers into the cluster. 

\paragraph*{Analysis of friction coefficient.}
In the simulations to measure the effective friction coefficient of the PomXY cluster, all PomZ dimers in the system are bound to the PomXY cluster and they cannot detach from it ($k_h = 0$) such that the number of cluster-bound PomZ dimers is constant. An external force is applied to the cluster and the force-velocity curve is recorded. More specifically, at least three different forces (\SI{0.005}{\pico \N}, \SI{0.01}{\pico \N}, \SI{0.02}{\pico \N}) are applied, and the average steady-state cluster velocity is calculated based on $100$ trajectories. Plotting the force against the velocity yields a linear dependence, and the friction coefficient can be obtained from the slope. In these simulations an infinitely extended PomXY cluster and nucleoid is considered, i.e.\ boundaries are neglected. This is done because otherwise the PomZ dimers would accumulate at one of the cluster ends.

\paragraph*{Analysis of oscillatory properties.}
In the simulations set up to study the oscillatory behavior of the cluster, the PomXY cluster starts at midnucleoid and its position is recorded over a long time (at least \SI{1000}{\min}). Initially, all PomZ dimers are in the cytosol, but the cluster movement only starts after $t_\text{min} = \SI{10}{\min}$, such that the PomZ dynamics can approach its stationary distribution. Two observables are of interest: the cluster position distributions and the Fourier spectrum of the cluster trajectories. In the case of the first, the histogram depicting the cluster positions of all runs is smoothed using a Gaussian moving average and peaks are identified in the smoothed profile, which are local maxima or minima. Depending on the parameters chosen, there might be no local minima. In this case, the cluster position distribution has a monomodal shape. If there is a minimum and the difference between the maximal and minimal peak is larger than $2\%$ of the maximal count, the profile is classified as bimodal. 

To determine if the cluster trajectories are oscillatory or not and to estimate the frequencies of cluster oscillations, the procedure used is as follows: For each run, the temporal average of the cluster position is subtracted from the cluster trajectory and a fast Fourier transform of the resulting data is performed. The modulus of the Fourier-transformed cluster position for each run is summed, and the resulting spectrum is smoothed using a moving average with Gaussian weights. Then the largest peak is identified in the smoothed data with a minimal peak height $10\%$ higher than the value corresponding to the smallest frequency, $f_\text{min} = 1/T_\text{max}$, in the smoothed data set ($T_\text{max}$ is the duration of the signal considered in the Fourier transformation). If there is a peak, the cluster trajectory is oscillatory with the frequency determined by the peak in the Fourier spectrum. On the other hand, if no peak is found, the trajectories are classified as ``non-oscillatory''.

\subsection*{Stationary cluster simulations}
Simulations with a fixed position of the PomXY cluster are performed to measure the force exerted by a single PomZ dimer on the cluster (``one-particle simulations'') or to measure the PomZ dimer flux into the cluster and the forces exerted on the cluster for an arbitrary number of PomZ dimers in the system. In these simulations, the PomZ dimer(s) are initially in the cytosol. Typically, in the simulations with a dynamic PomXY cluster, the number of cluster-bound PomZ dimers is large. Together with a large friction coefficient of the PomXY cluster, this results in only small movements of the cluster. The stationary cluster simulations can be used as an approximation in these cases.  

\paragraph*{One-particle simulations.}
To determine the force typically exerted by a single PomZ dimer on the PomXY cluster, simulations with only one PomZ dimer in the system are performed. Here, the PomXY cluster is located far away from the nucleoid boundaries (at midnucleoid) and the PomZ dimer attaches to a lattice site on the right side of the cluster that is so far away from the cluster that no interaction with the cluster is possible. In the simulations, we record the nucleoid and cluster binding site positions when the PomZ dimer attaches to the PomXY cluster, and the force exerted on the PomXY cluster integrated over time and averaged over time for a number, $N_\text{runs}$, of PomZ dimers interacting with the PomXY cluster. To obtain the constant $C = f/k_h$, the ensemble average of the time-averaged force, $f$, needs to be determined. This quantity is calculated as follows:
\begin{equation*}
f = \frac{\sum_i f_i t_i}{\sum_i t_i} = \frac{\sum_i f_i^\text{int}}{\sum_i t_i},
\end{equation*}
with $f_i$ and $f_i^\text{int}$ the time-averaged and time-integrated force exerted by a single PomZ dimer interacting with the PomXY cluster and $t_i$ the corresponding time of interaction. In this definition of $f$, each time-averaged force is weighted by the time the particle remains attached to the PomXY cluster when calculating the mean. Note that the constant $C$ can also be expressed in terms of the ensemble average of the time-integrated force, $f^\text{int}$:
\begin{equation*}
C = \frac{f}{k_h} = \frac{\sum_i f_i^\text{int}}{k_h \sum_i t_i} \approx \frac{\sum_i f_i^\text{int}}{N_\text{runs}} = f^\text{int},
\end{equation*}
because $\frac{1}{N_\text{runs}} \sum_i t_i = \frac{1}{k_h}$ for large $N_\text{runs}$.

\begin{acknowledgments}
We thank Dominik Schumacher, Lotte S\o gaard-Andersen, Jean-Charles Walter, Andrea Parmeggiani, Matthias Kober, Isabella Graf, Johannes Knebel, Emanuel Reithmann, Karl Wienand and Jacob Halatek for helpful discussions. This research was supported by a DFG fellowship through the Graduate School of Quantitative Biosciences Munich, QBM (SB), the Deutsche Forschungsgemeinschaft (DFG) via project P03 within the Transregio Collaborative Research Center (TRR 174) ``Spatiotemporal Dynamics of Bacterial Cells'' (SB, EF), and the German Excellence Initiative via the program ``Nanosystems Initiative Munich'' (EF).
\end{acknowledgments}


\cleardoublepage

\newpage

\widetext

\begin{center}
\textbf{\large Supplemental Material: Regulation of Pom cluster dynamics in \textit{Myxococcus xanthus}}
\end{center}
\setcounter{equation}{0}
\setcounter{figure}{0}
\setcounter{table}{0}
\setcounter{page}{1}
\makeatletter
\renewcommand{\theequation}{S\arabic{equation}}
\renewcommand{\thefigure}{S\arabic{figure}}
\renewcommand{\bibnumfmt}[1]{[S#1]}

\section{\label{S1_File}Stationary solution of the RD model.}
To get analytical expressions for the flux difference of PomZ dimers into the cluster and the number of PomZ dimers bound to the PomXY cluster, we described the PomZ dynamics in terms of reaction-diffusion equations (Eq.~\ref{eq:RD_model1}-\ref{eq:RD_model6} in the main text). We are interested in the steady-state solutions. Eq.~\ref{eq:RD_model1} and \ref{eq:RD_model2} is solved by imposing no flux boundary conditions at $x=0$ and $x=L$ and setting the values for the concentration, $c(x,t)$, and its derivative, $\partial_x c(x,t)$, inside and outside of the PomXY cluster region equal at the cluster boundaries, $x=x_c \pm L_c/2$. The second condition, an equivalence of the first derivative of the concentration, is due to balance of diffusive fluxes at $x = x_c \pm L_c/2$. Since the diffusion constants of PomZ dimers bound to the nucleoid, but not bound to the PomXY cluster, are the same below the PomXY cluster and away from it, the first derivative has to be set equal at $x = x_c \pm L_c/2$. The resulting expression for $c(x,t)$ can be used to solve Eq.~\ref{eq:RD_model3} in the steady state. For the cluster-bound PomZ dimers, $c_b(x,t)$, no-flux boundary conditions hold at the PomXY cluster's edges. The last unknown is $N_\text{cyto}$, which is determined by mass conservation, 
\begin{equation}
\int_0^L c(x)\, dx + \int_{x_c-L_c/2}^{x_c+L_c/2} c_b(x) \, dx + N_\text{cyto} = N_\text{total} 
\end{equation}
or equivalently by solving Eq.~\ref{eq:RD_model4} in the steady state. The solutions for $c(x), c_b(x)$ and $N_\text{cyto}$ are obtained with \textit{Mathematica}. They are quite lengthy and hence not written out here explicitly. The flux difference of PomZ dimers into the cluster can then be calculated as follows:
\begin{equation}
j_\text{diff} = D_\text{nuc} \partial_x c(x)|_{x_c+L_c/2} + D_\text{nuc} \partial_x c(x)|_{x_c-L_c/2}.
\end{equation}
We chose the sign of the flux difference to be positive if more PomZ dimers arrive from the right than from the left side. The number of PomZ dimers bound to the PomXY cluster is given by:
\begin{equation}
N(x_c) = \int_{x_c-L_c/2}^{x_c+L_c/2} c_b(x) dx.
\end{equation}

\section{\label{S2_File}Derivation of the effective friction coefficient of the PomXY cluster.}
\begin{figure}[h!]
\includegraphics[]{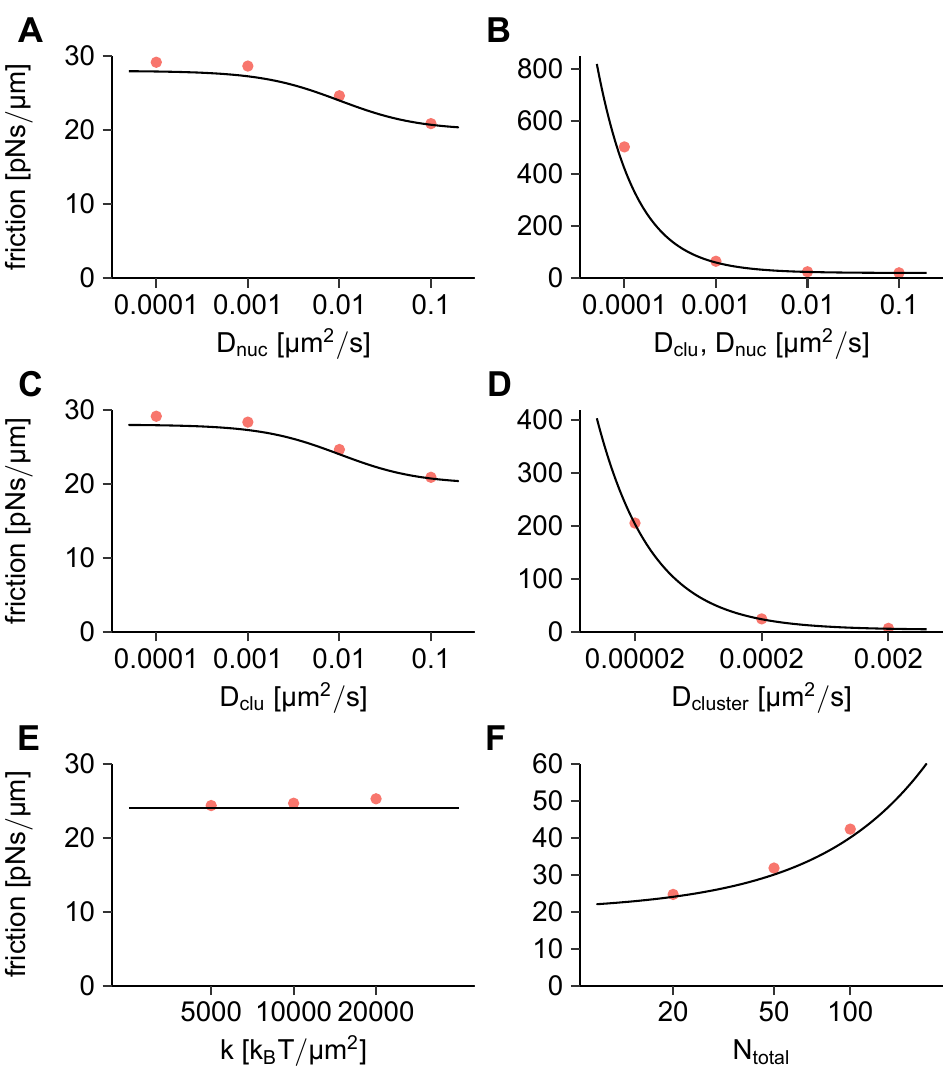}
\caption{\textbf{Friction coefficient of the PomXY cluster.} (A-F) We determined the friction coefficient of the PomXY cluster with $N = 20$ PomZ dimers bound to it, when the diffusion constant of PomZ on the nucleoid and the PomXY cluster (A-C), the cytosolic diffusion constant of the PomXY cluster (D), and the spring stiffness of the PomZ dimers (E) is varied. Finally, we varied the PomZ dimer number bound to the PomXY cluster keeping all other parameters fixed (F). In all cases, the friction coefficients obtained from simulations (red dots) agree with the theoretical prediction (black line, Eq.~\ref{eq:effective_friction}). The effective friction coefficient of the PomXY cluster increases with an increasing friction of PomZ on the nucleoid and the PomXY cluster, an increasing cytosolic cluster friction and an increasing cluster-bound PomZ dimer number. It does not depend on the spring stiffness of the PomZ dimers for the parameter range considered. For more details see the Materials and methods section. In the simulations performed for this Figure, the nucleoid and PomXY cluster are infinitely extended, all PomZ dimers in the system are bound to the cluster, the ATP hydrolysis rate is set to zero and the other parameters are as in Table \ref{S1_Table} if not explicitly given.}
\label{S6_Fig}
\end{figure}

Our aim is to get an analytical expression for the effective friction coefficient of the PomXY cluster, i.e.\ the friction coefficient when the cluster is tethered to the nucleoid by $N$ PomZ dimers. We consider an infinitely extended PomXY cluster and nucleoid to exclude boundary effects. Note that the absolute positions of the PomZ dimers do not matter, only the difference between the positions of the nucleoid and cluster binding sites matters. Hence, all PomZ dimers can be moved to the same nucleoid position. The position of the cluster binding site then has a distribution that is peaked at the position of the nucleoid binding site if no force is exerted to the PomXY cluster and the peak of the distribution is shifted to the right if the PomXY cluster is pulled to the right by an external force. Let us denote the position of the cluster by $x(t)$. The position of the nucleoid binding site of all PomZ dimers is denoted $x^\text{nuc}(t)$, and the average position of the cluster binding sites $x(t) + \Delta x^\text{clu}(t)$. Then the following equations hold 
\begin{align}
\partial_t x(t) &= \frac{F}{\gamma_c} - N \frac{k}{\gamma_c}(x(t) + \Delta x^\text{clu}(t) - x^\text{nuc}(t)),\\
\partial_t\Delta x^\text{clu}(t) &= -\frac k \gamma_\text{clu}(x(t) + \Delta x^\text{clu}(t) - x^\text{nuc}(t)),\\
\partial_t x^\text{nuc}(t) &= -\frac k \gamma_\text{nuc}(-x(t) - \Delta x^\text{clu}(t) + x^\text{nuc}(t)),
\end{align} 
with $\gamma_\text{clu}$, $\gamma_\text{nuc}$ the friction coefficient of PomZ dimers on the PomXY cluster and the nucleoid, respectively. They are related to the diffusion constants of a PomZ dimer on the nucleoid and PomXY cluster via Stokes-Einstein, $D_{\text{clu}/\text{nuc}} = k_BT/\gamma_{\text{clu}/\text{nuc}}$. 
We solved the coupled ODE system above using \textit{Mathematica}. Taking the time derivative of the position of the PomXY cluster, $x(t)$, and dividing $F$ by this expression for the velocity of the cluster yields the effective friction coefficient of the PomXY cluster in dependence of $N$ and the other model parameters:
\begin{equation}
\gamma(t, N) = \frac{e^{k(1/\gamma_\text{clu}+1/\gamma_\text{nuc} + N/\gamma_c)t}\gamma_c (\gamma_c (\gamma_\text{clu} + \gamma_\text{nuc}) + \gamma_\text{clu} \gamma_\text{nuc} N)}{e^{k(1/\gamma_\text{clu}+1/\gamma_\text{nuc} + N/\gamma_c)t}\gamma_c (\gamma_\text{clu} + \gamma_\text{nuc}) + \gamma_\text{clu} \gamma_\text{nuc} N}.
\end{equation}
For large times ($t \rightarrow \infty$), this simplifies to
\begin{equation}
\gamma (N) = \gamma_c + \frac{\gamma_\text{clu}\gamma_\text{nuc} N}{\gamma_\text{clu} + \gamma_\text{nuc}}.
\label{eq:effective_friction_SI}
\end{equation}
We find that the effective friction coefficient of the cluster is given by the cytosolic friction plus an additional term that increases linearly with the number of PomZ dimers bound to the cluster, $N$. This expression is a generalization of the term derived by Lansky et al., \cite{Lansky2015} using a force-balance argument. The simulations fit well with this theoretical curve if we use an infinitely extended nucleoid and PomXY cluster, i.e.\ we neglect the boundary conditions (Fig.~\ref{S6_Fig}). If the nucleoid and PomXY cluster size is finite in the simulations, the measured effective friction coefficient is higher than the analytical result derived here, because the PomZ dimers move to the rear cluster's edge when the cluster is pulled forward and the reflecting boundary conditions of the cluster for the movement of PomZ's cluster binding site increase the friction. However, in the dynamic cluster simulations, i.e.\ simulations with a finite nucleoid and PomXY cluster, which is moved on the nucleoid by PomZ dimer interactions, cluster-bound PomZ dimers can detach into the cytosol with the ATP hydrolysis rate $k_h$. Hence, the effect of accumulation of PomZ dimers at the rear cluster's edge is diminished. In these simulations, the number of cluster-bound PomZ dimers depends on the position of the cluster on the nucleoid. In this case, we replace Eq.~\ref{eq:effective_friction_SI} with
\begin{equation}
\gamma (x_c) = \gamma_c + \frac{\gamma_\text{clu}\gamma_\text{nuc} N(x_c)}{\gamma_\text{clu} + \gamma_\text{nuc}}.
\end{equation}

\begin{table}[!ht]
\begin{tabular}{| l | l | l | l |}
\hline
{\bf Parameter} & {\bf Symbol} & {\bf Value} \\ \thickhline
Number of PomZ dimers & $N_\text{total}$ & 100$^*$ \\ \hline
Length of nucleoid & $L$ & \SI{5}{\um}$^*$ \\ \hline
Length of cluster & $L_c$ & \SI{0.7}{\um} \\ \hline
Attachment rate PomZ to nucleoid & $k_\text{on}$ & \SI{0.1}{\per\second} \\ \hline
Attachment rate PomZ to PomXY cluster (unstretched) & $k_a^0$ & \SI{500}{\per\second\per\um} \\ \hline 
Diffusion constant PomZ on nucleoid & $D_\text{nuc}$ & \SI{0.01}{\um^2\per\second}  \\ \hline
Diffusion constant PomZ on PomXY cluster & $D_\text{clu}$ & \SI{0.01}{\um^2/s} \\ \hline
ATP hydrolysis rate PomZ bound to PomXY cluster & $k_h$ & \SI{0.01}{\per\second} \\ \hline
Diffusion constant PomXY cluster in cytosol & $D_\text{cluster} = {k_B T}/{\gamma_c}$ & \SI{0.0002}{\um^2/s} \\ \hline
Spring constant PomZ dimers & $k$ & \SI{e4}{k_B T/\um^2} \\ \hline 
Lattice spacing & $a$ & \SI{0.01}{\um} \\ \hline
\end{tabular}
\begin{flushleft} 
$^*$ In the one-particle simulations, the total number of PomZ dimers, $N_\text{total}$, is one and the nucleoid length is chosen to be $L = \SI{2.1}{\um}$ to decrease computation time. 
\end{flushleft}
\caption{{\bf Parameters used in the simulations.} If not explicitly stated otherwise the values for the model parameters shown here are those used in the simulations. The values for the total number of PomZ dimers, $N_\text{total}$, the length of the nucleoid and the cluster, $L$ and $L_c$, and the ATP hydrolysis rate of PomZ dimers interacting with the PomXY cluster and DNA, $k_h$, are determined from experiments in \textit{M. xanthus} cells \cite{Schumacher2017a}. The attachment rate of PomZ to the nucleoid, $k_\text{on}$, was chosen to a value in between those given in \cite{Ietswaart2014} (\SI{50}{s^{-1}}) and \cite{Lim2014} (\SI{0.03}{s^{-1}}) for ParA attachment to the nucleoid. The cluster dynamics is only slightly affected when this parameter is varied over a broad parameter range (see Fig.~\ref{fig:parameter_sweep}A). The same holds true for the attachment rate of PomZ to the PomXY cluster (see Fig.~\ref{fig:parameter_sweep}B). The diffusion constant of a PomZ dimer on the nucleoid is approximated by the diffusion constant of DNA-bound ParA dimers as given in \cite{Lim2014}, and typically the same value was chosen for the diffusion constant of PomZ on the PomXY cluster in the simulations. For the diffusion constant of the PomXY cluster in the cytosol, i.e.\ not tethered to the nucleoid via PomZ dimers, we used a value similar to those for a plasmid \cite{Ietswaart2014}, \SI{0.0003}{\um^2/s} and a partitioning complex \cite{Lim2014}, \SI{0.0001}{\um^2/s}. The effective spring stiffness, $k$, in our model, which accounts for the elasticity of the nucleoid and the PomZ dimers, is approximated by the value for the stiffness of the bond between the plasmid and the nucleoid DNA via ParA dimers used in \cite{Hu2017a}. The lattice spacing $a$ was chosen such that feasible computation times of our simulations are achieved. To test our choice of the lattice spacing, we performed simulations with $a=\SI{0.005}{\um}$ and did not observe remarkable changes in the cluster dynamics.}
\label{S1_Table}
\end{table}

\begin{figure}[h!]
\includegraphics[]{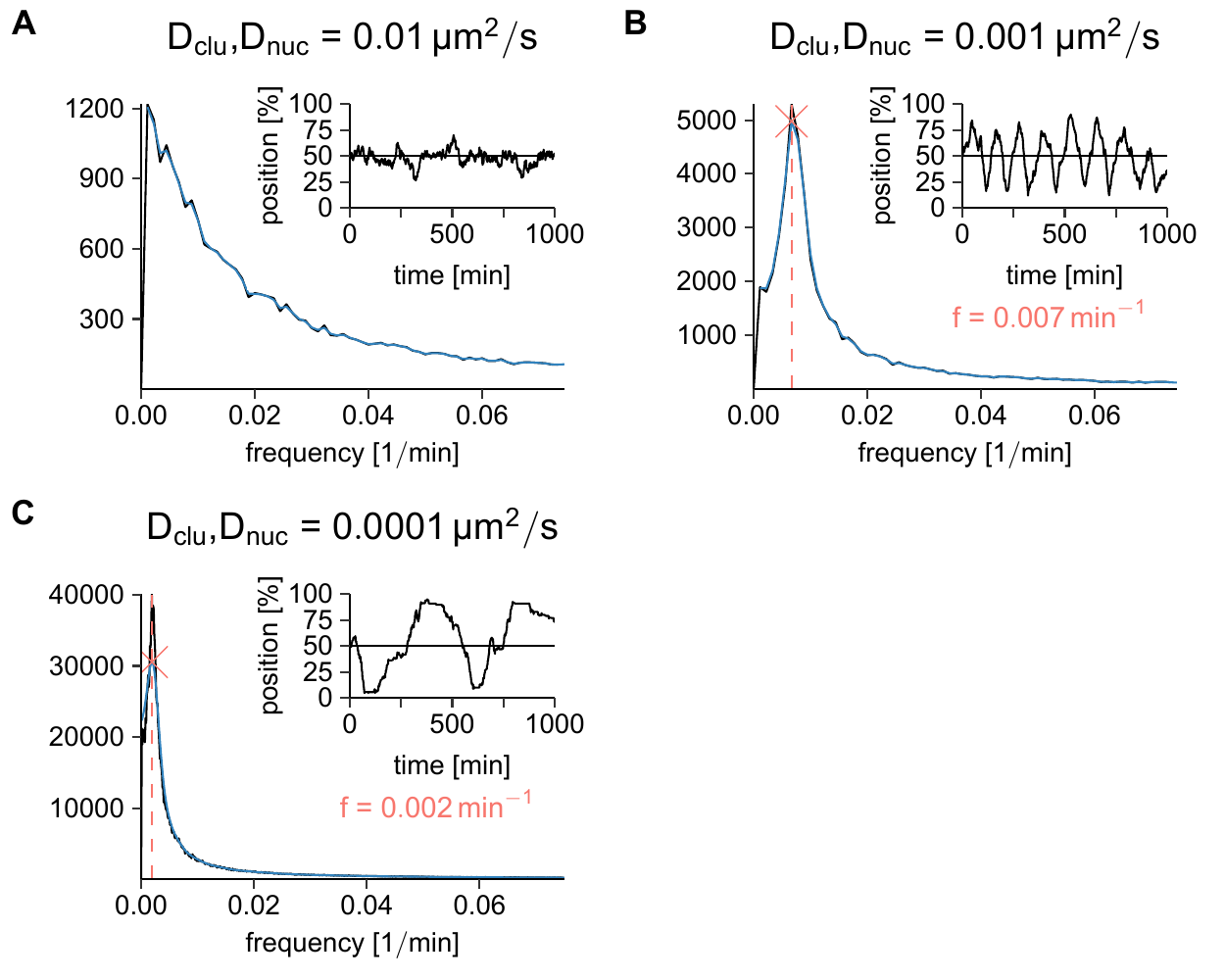}
\caption{\textbf{Midnucleoid localization vs. oscillatory movements.} (A-C) The average magnitude of the fast Fourier transform signal (black line) is smoothed using a moving average with Gaussian weights (blue line) to determine whether there is a peak in the Fourier spectrum or not (for details see Materials and methods). The insets show a cluster trajectory for one run. The diffusion constants of PomZ on the nucleoid and PomXY cluster are varied over two orders of magnitudes; the other parameters are chosen as in Table \ref{S1_Table}. For the Fourier analysis we performed $100$ runs of the simulation for $\ge$ \SI{1000}{\min} with a cluster starting at midnucleoid.}
\label{S1_Fig}
\end{figure}

\begin{figure}[h!]
\includegraphics[]{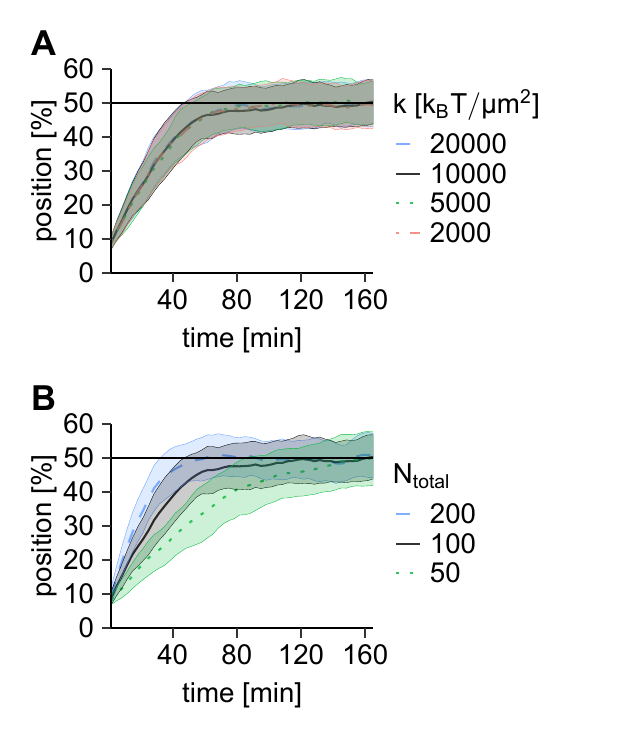}
\caption{\textbf{Additional parameter sweeps.} Same as in Fig.~\ref{fig:parameter_sweep}, but here we vary the spring stiffness, $k$ (A), and the total number of PomZ dimers, $N_\text{total}$ (B). The spring stiffness can be changed over an order of magnitude without changing the cluster dynamics. However, note that the attachment rate of PomZ dimers to the PomXY cluster is defined in such a way that the total attachment rate to the cluster depends on $k$. The more PomZ dimers are in the system, the faster the clusters move towards midnucleoid.}
\label{S2_Fig}
\end{figure}

\begin{figure}[h!]
\includegraphics[]{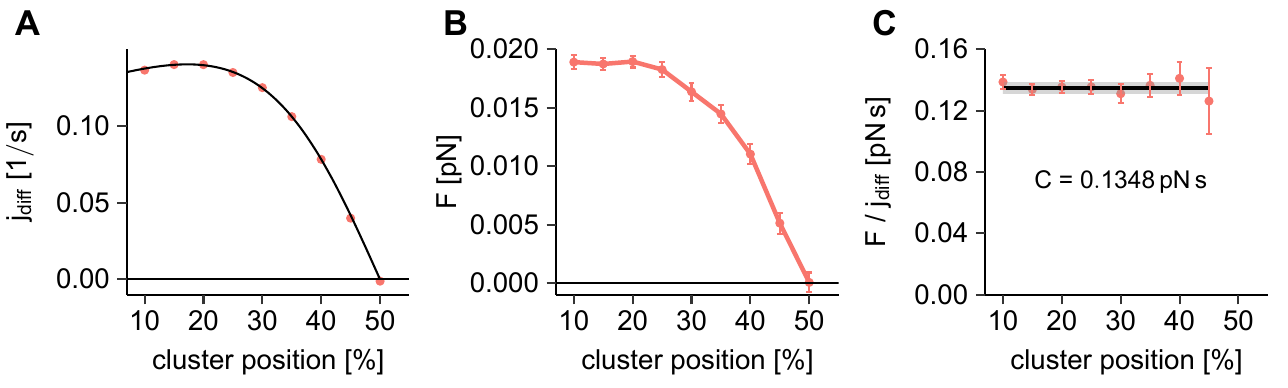}
\caption{\textbf{The net force is proportional to the flux difference for a stationary PomXY cluster.} We simulated the PomZ dynamics for a cluster that is kept fixed at different positions on the nucleoid. (A) The PomZ flux difference into the cluster, $j_\text{diff}$, obtained from the simulations (in red) agrees nicely with the predicted flux difference from the RD model (black line). (B) In the simulations, the total force exerted by the PomZ dimers on the PomXY cluster averaged over time, $F$, also decreases towards zero when the cluster moves from an off-center position towards midnucleoid. (C) The ratio of the total force and the PomZ flux difference (red dots) does not change remarkably with the cluster position, as expected. We discard the value at $50\%$ nucleoid length, because both the flux difference and the total force are supposed to be zero in this case. The black line is a fit of a constant curve to the data with fit parameter $C = F/j_\text{diff} = \SI{0.1348}{\pico\N/\s}$. The grey area indicates the $95\%$ confidence interval of the fit. The simulated values for the flux difference and the total force are obtained by averaging over $10$ realisations of the stochastic simulation per cluster position (the error bars show the $95\%$ confidence interval). The simulation parameters are as in Table \ref{S1_Table}.}
\label{S3_Fig}
\end{figure}

\begin{figure}[h!]
\includegraphics[]{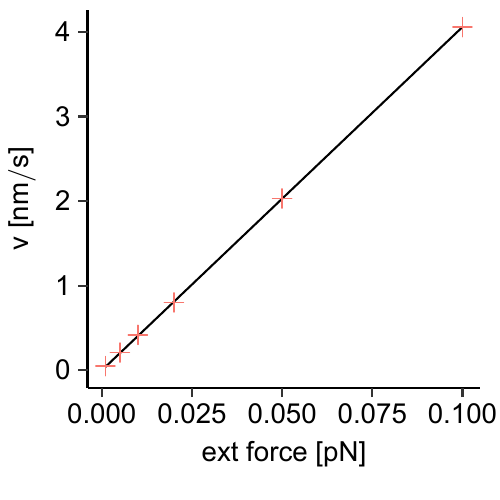}
\caption{\textbf{Force-velocity curve.} The average velocity of the PomXY cluster increases linearly with the external force applied to the cluster. For different external force values we simulated $100$ trajectories of a PomXY cluster and determined the average steady-state velocity of the cluster (red crosses). A linear fit to the data (black line) matches the simulation results well and yields the effective friction coefficient of the cluster, which is the inverse of the slope. In the simulations an infinitely extended cluster and nucleoid was used (for details see Materials and methods). We simulated $N_\text{total} = 20$ PomZ dimers, all bound to the PomXY cluster, and the ATP hydrolysis rate $k_h$ was set to zero. The other parameters are as in Table \ref{S1_Table}.}
\label{S4_Fig}
\end{figure}

\begin{figure}[h!]
\includegraphics[]{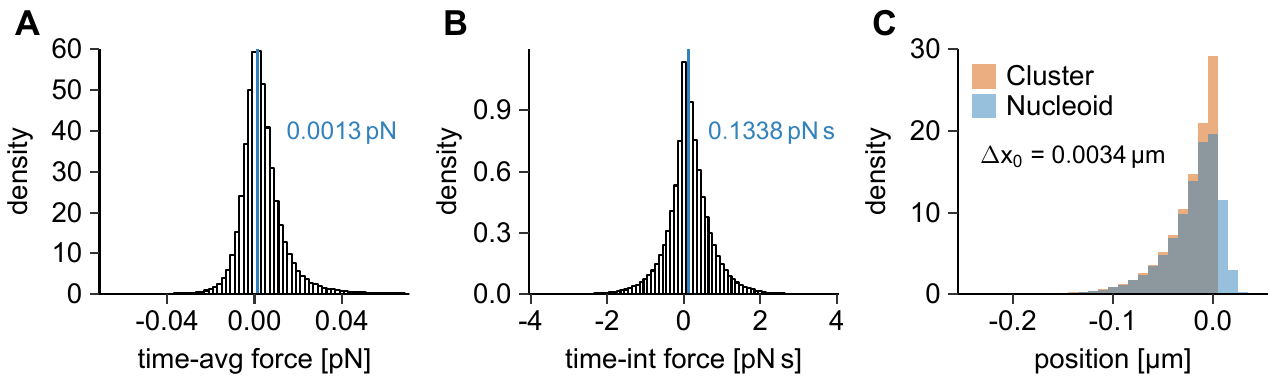}
\caption{\textbf{Single particle force generation.} To determine the constant $C$, simulations with only one PomZ dimer and a fixed PomXY cluster position are performed (parameters as in Table \ref{S1_Table}). The PomZ dimer stochastically attaches to the rightmost side of the nucleoid, diffuses on the nucleoid, interacts with the PomXY cluster and then detaches from the PomXY cluster and the nucleoid. We simulated more than $\num{400000}$ particle-cluster interactions and recorded the distributions of time-averaged forces (A), time-integrated forces (B) and the distributions of the binding sites of the PomZ dimers on the nucleoid and cluster when attaching to the PomXY cluster (C). 
The ensemble average of the time-averaged force, weighting each time-averaged force with the corresponding time a PomZ dimer is attached to the cluster, is positive $f=\SI[separate-uncertainty]{0.1335(00009)e-2}{\pico \N}$ (the error is the standard error of the mean). The same holds true for the mean time-integrated force $f_\text{int}=\SI[separate-uncertainty]{0.1338(00009)}{\pico \N \s}$, which implies that a PomZ dimer arriving at the cluster from the right on average exerts a net force to the right. When attaching to the PomXY cluster, PomZ dimers are typically slightly stretched towards the PomXY cluster, which yields an average distance between the nucleoid and cluster binding site of $\Delta x_0 = \SI{0.0034}{\um}$.}
\label{S5_Fig}
\end{figure}

\begin{figure}[h!]
\includegraphics[]{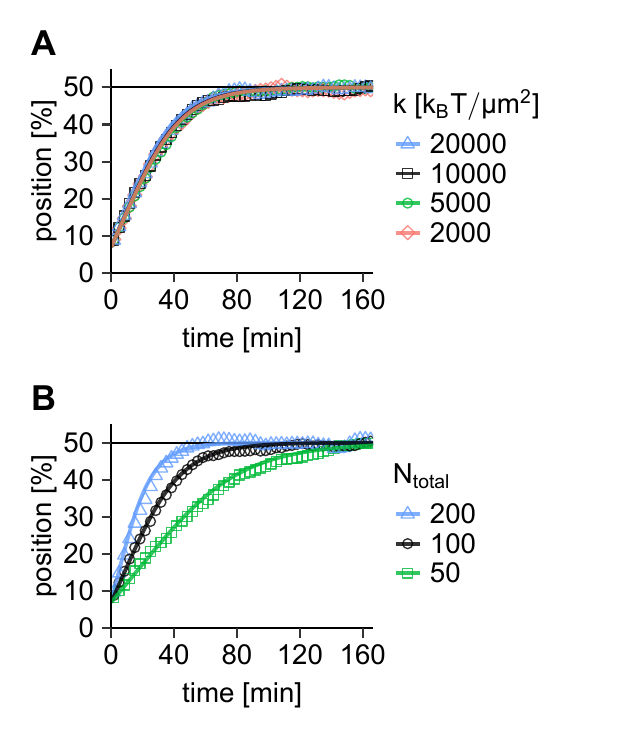}
\caption{\textbf{Comparison of the average cluster trajectory from simulations and our semi-analytical approximation for additional parameters.} Same as in Fig.~\ref{fig:average_cluster_trajectory}, when the spring stiffness $k$ (A) and the total PomZ dimer number $N_\text{total}$ (B) is varied. The average cluster trajectories are the same as shown in Fig.~\ref{S2_Fig}.}
\label{S7_Fig}
\end{figure}

\begin{figure}[h!]
\includegraphics[]{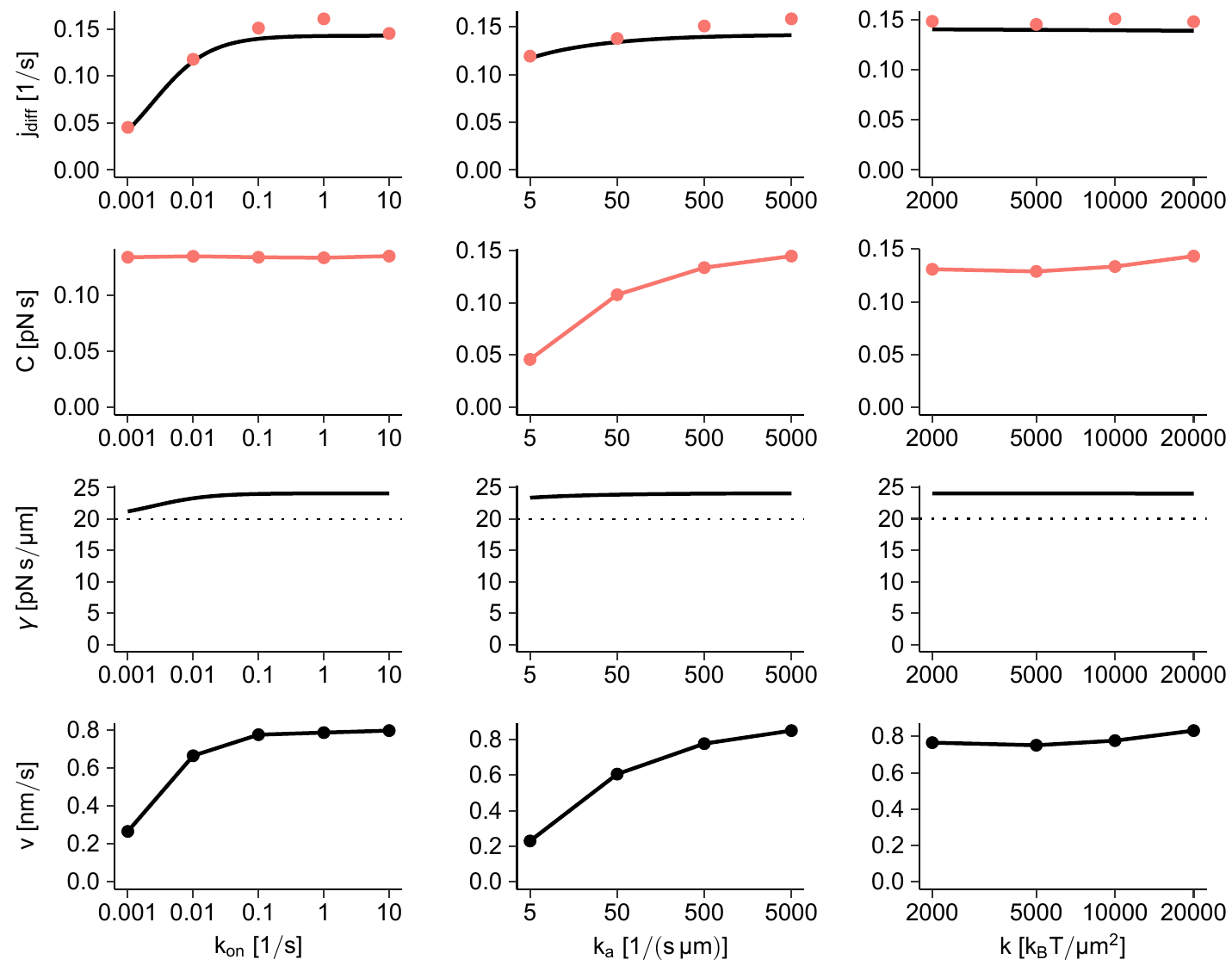}
\caption{\textbf{Force generation results for additional parameters.} Same as in Fig.~\ref{fig:force_generation} for parameter sweeps varying the attachment rate to the nucleoid $k_\text{on}$, the attachment rate to the PomXY cluster $k_a$ and the spring stiffness $k$. An increase in $k_\text{on}$ and $k_a$ increases the velocity of the cluster towards midnucleoid. An increase in $k$ leads to stiffer springs and hence less stretched PomZ dimers, but on the other hand, the force, which is linear in $k$, is increased. This results in a more or less constant value for $C$ and also a constant velocity of the cluster when varying $k$ over one order of magnitude. Note that a change in the spring stiffness also changes the total attachment rate of PomZ dimers to the PomXY cluster.}
\label{S8_Fig}
\end{figure}

\begin{figure}[h!]
\includegraphics[]{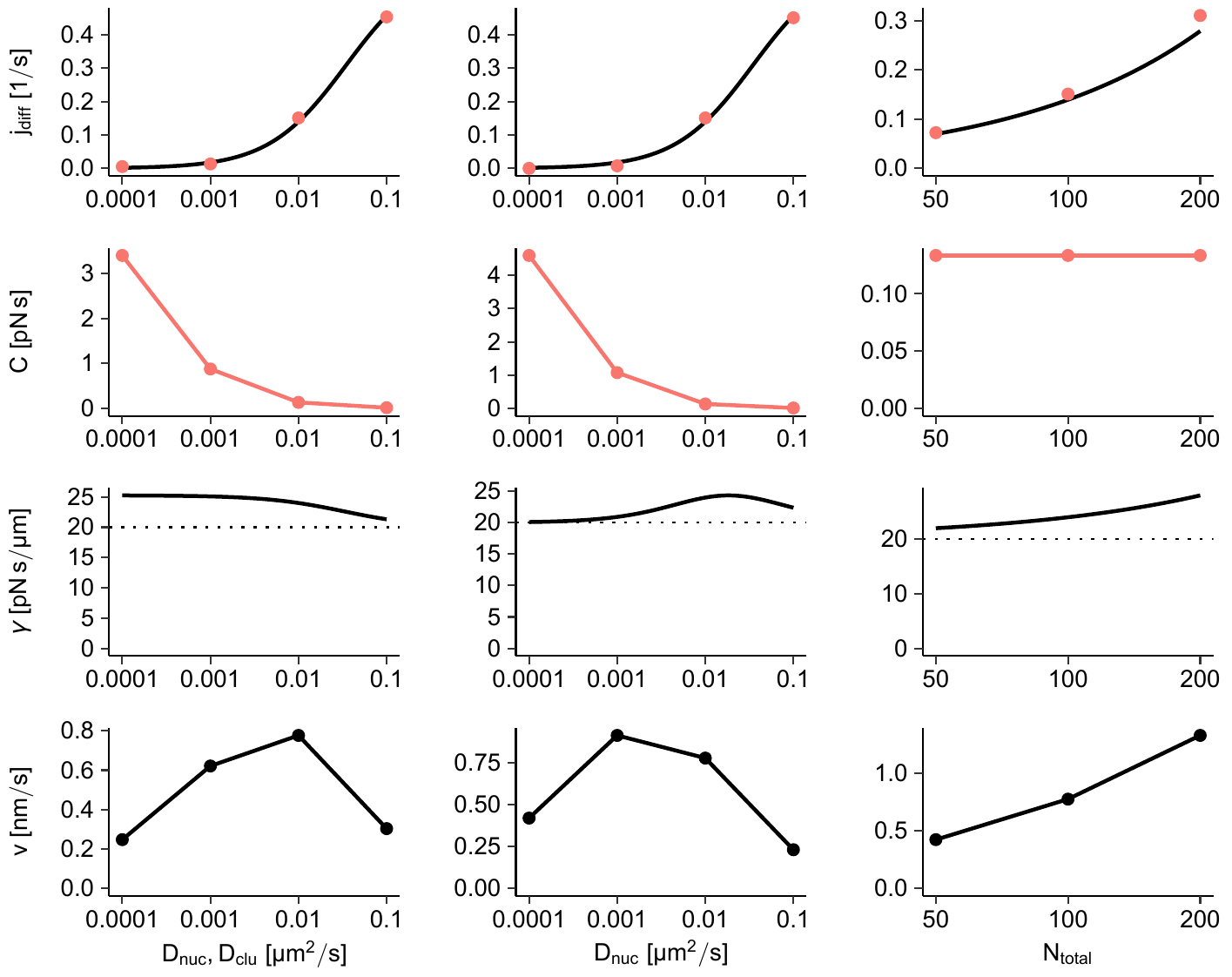}
\caption{\textbf{Force generation results for additional parameters.} Same as in Fig.~\ref{fig:force_generation} for parameter sweeps varying the diffusion constants of PomZ on the PomXY cluster and the nucleoid ($D_\text{nuc} = D_\text{clu}$), the diffusion constant of PomZ on the nucleoid, $D_\text{nuc}$, and the total PomZ dimer number, $N_\text{total}$. For very small diffusion constants of PomZ on the nucleoid our semi-analytical approach breaks down (see Fig.~\ref{fig:average_cluster_trajectory}). An increase in the total PomZ dimer number increases the PomZ flux difference into the cluster, but does not change the constant $C$, since $C$ is an observable for a single particle. With an increased number of PomZ dimers, the number of cluster-bound PomZ dimers also increases, which in turn leads to a higher effective friction coefficient. Nevertheless, the velocity of the cluster, which is proportional to the flux difference and inversely proportional to the effective friction coefficient, increases with the PomZ dimer number.}
\label{S9_Fig}
\end{figure}

\begin{figure}[h!]
\includegraphics[]{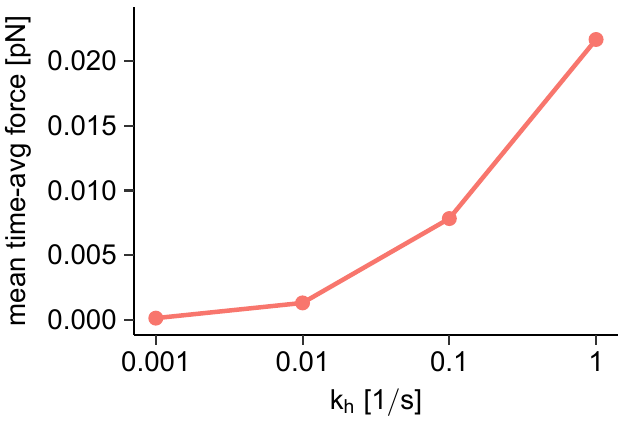}
\caption{\textbf{Mean time-averaged force for different $k_h$ values.} The ensemble average of the time-averaged force a single particle exerts on the PomXY cluster increases with the hydrolysis rate $k_h$. The larger the hydrolysis rate, the shorter the interaction time of the PomZ dimer with the PomXY cluster. Since the PomZ dimers typically attach close to the cluster's edge and over time diffuse towards the center of the cluster, the average force exerted by the particle decreases over time. Therefore, a shorter interaction time yields a larger time-averaged force. If not explicitly given in the Figure, the parameters are as in Table \ref{S1_Table}.}
\label{S10_Fig}
\end{figure}

\begin{figure}[h!]
\includegraphics[]{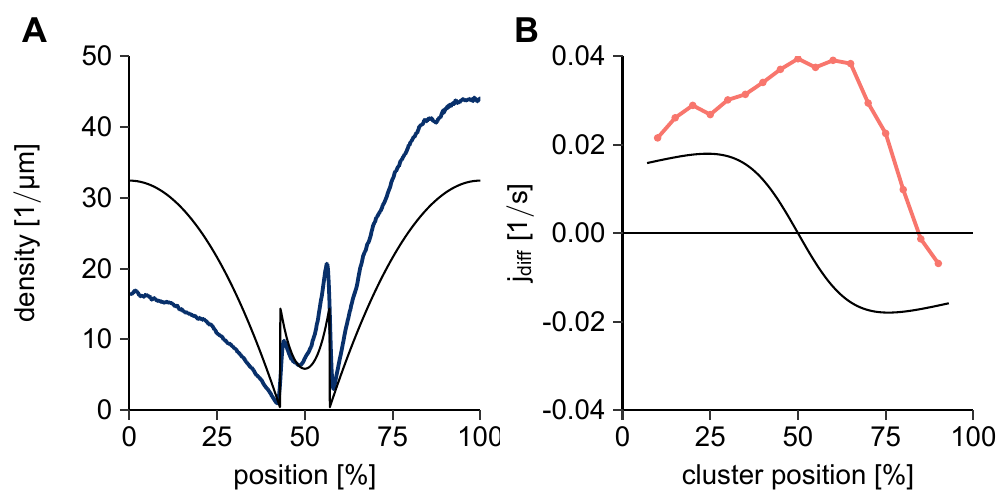}
\caption{\textbf{PomZ density and flux for an oscillatory cluster.} PomZ density along the nucleoid (A) and PomZ flux difference into the cluster (B) as shown in Fig.~\ref{fig:comparison_rd_stochsim} using the parameters in Table \ref{S1_Table}, but a reduced diffusion constant of PomZ on the nucleoid and PomXY cluster ($D_\text{nuc} = D_\text{clu} = \SI{0.001}{\um\squared\per\second}$).}
\label{S11_Fig}
\end{figure}

\begin{figure}[h!]
\includegraphics[]{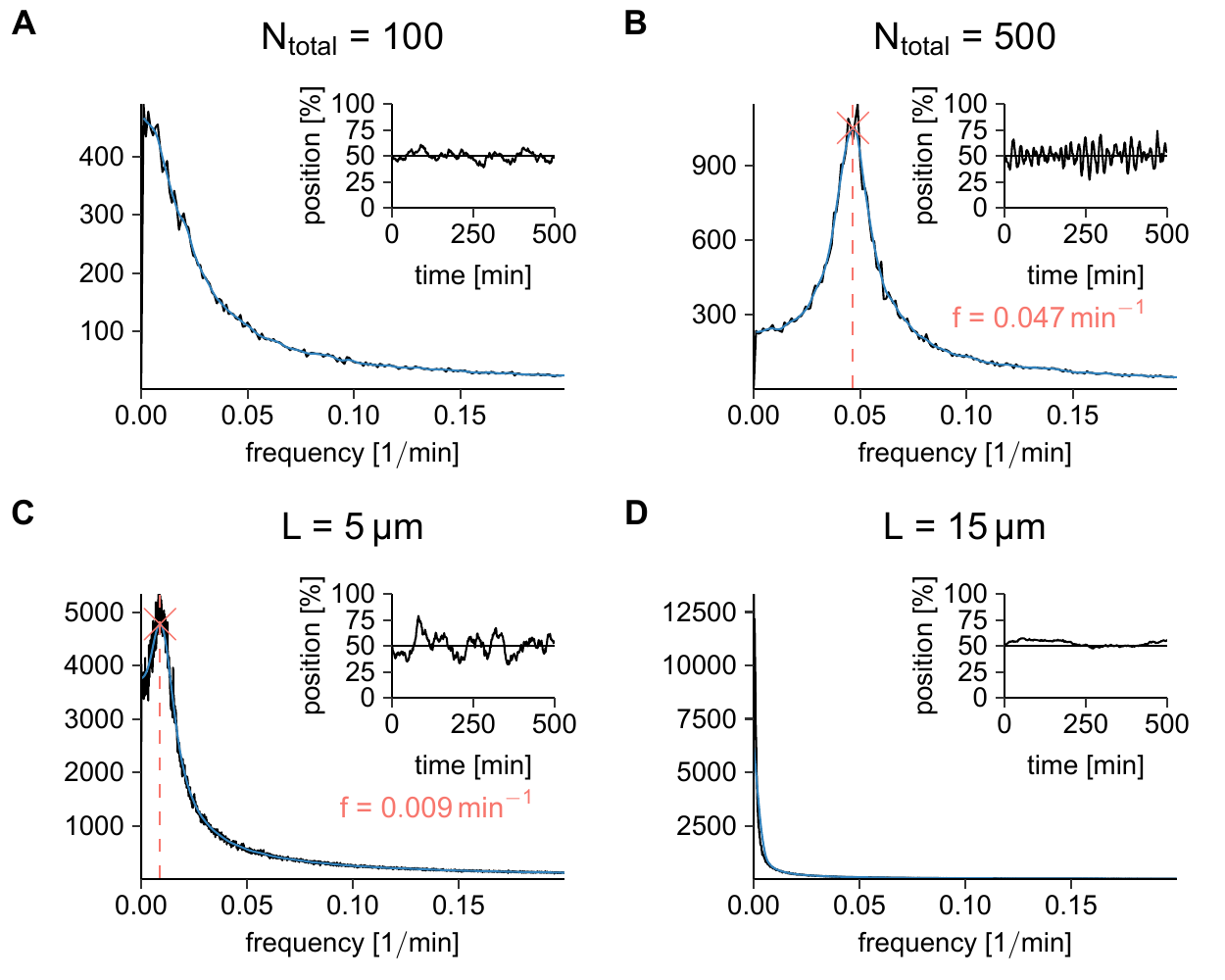}
\caption{\textbf{Frequency analysis of the cluster dynamics varying $N_\text{total}$ and $L$.} The averaged fast fourier transform of the cluster trajectories and a single trajectory (inset) are shown (see Fig.~\ref{S1_Fig} and Materials and methods for details). (A, B) When the total PomZ dimer number is increased from $N_\text{total} = 100$ to $N_\text{total} = 500$, the cluster dynamics change from fluctuating around midnucleoid to oscillatory with a frequency of $f = \SI{0.047}{\min^{-1}}$ ($k_h = \SI{0.1}{\s^{-1}}$, other parameters as in Table \ref{S1_Table}). For the Fourier analysis we performed $100$ runs of the simulation for \SI{1000}{\min} with a cluster starting at midnucleoid. (C, D) When the nucleoid length, $L$, is increased from $L = \SI{5}{\um}$ to $L = \SI{15}{\um}$, the peak in the Fourier spectrum, which indicates on average oscillations of the clusters with a frequency $f = \SI{0.009}{min^{-1}}$, disappears ($D_\text{nuc}, D_\text{clu} = \SI{0.003}{\um^2/s}$, other parameters as in Table \ref{S1_Table}). We performed $100$ runs of the simulation for \SI{10000}{\min} with a cluster starting at midnucleoid.}
\label{S12_Fig}
\end{figure}

\end{document}